\documentclass[fleqn,usenatbib]{mnras}

\usepackage{newtxtext,newtxmath}

\usepackage[T1]{fontenc}

\DeclareRobustCommand{\VAN}[3]{#2}
\let\VANthebibliography\thebibliography
\def\thebibliography{\DeclareRobustCommand{\VAN}[3]{##3}\VANthebibliography}

\usepackage{graphicx}	
\usepackage{amsmath}	

\title[X-ray spectral study of 4U 1702-429]{Probing the accretion geometry of the atoll source 4U 1702-429 in different spectral states with \textit{NICER}, \textit{NuSTAR}, and \textit{AstroSat}}
\author[S. Banerjee et al.]{
Srimanta Banerjee,$^{1}$\thanks{E-mail: srimanta.banerjee4@gmail.com}
and Jeroen Homan$^{2}$
\\
$^{1}$Inter-University Centre for Astronomy and Astrophysics (IUCAA), PB No.4, Ganeshkhind, Pune-411007, India\\
$^{2}$Eureka Scientific, Inc., 2452 Delmer Street, Oakland, CA 94602, USA\\
}

\date{Accepted 2024 February 16. Received 2024 January 12; in original form 2023 September 26}

\pubyear{2024}

\begin{document}
\label{firstpage}
\pagerange{\pageref{firstpage}--\pageref{lastpage}}
\maketitle

\begin{abstract}
We perform a comprehensive spectral study of a carefully selected sample (total exposure $\sim 50.5$ ks) of \textit{NICER} observations of the atoll neutron star low-mass X-ray binary 4U 1702-429. Our sample encompasses nearly all classical spectral states found within the \textit{NICER} dataset. We require two thermal emission components, originating from the accretion disc and the boundary layer,
to describe the soft state spectra in the energy band 0.3-10.0 keV. In contrast, in our model, only the disc component directly contributes to the intermediate/hard state. Additionally, we use a thermally Comptonized component (or a power-law with pegged normalization) to represent the hard coronal emission in the soft and intermediate/hard state spectra. The boundary layer emerges as the principal source providing soft seed photons for Comptonization across all spectral states. In contrast to a previously held assertion regarding this source, our analyses reveal a decrease in the inner disc temperature coupled with the retreat of the inner disc from the neutron star surface as the source evolves from the soft to the intermediate/hard state.
The reflection features are either absent or weak ($\sim 3\sigma-4\sigma$) in all these observations. Further investigation using broad-band \textit{NuSTAR} (3.0-50.0 keV) and \textit{AstroSat} spectra (1.3-25.0 keV) shows a slightly stronger iron emission line ($\sim 5.8\sigma$) in the \textit{NuSTAR} spectra. 
However, this feature is not significantly detected in the \textit{AstroSat} observation. The \textit{AstroSat} data suggests a highly ionized disc, explaining the absence of reflection features. In the case of \textit{NuSTAR}, a truncated disc is likely responsible for the weak reflection features.
\end{abstract}


\section{Introduction}

A low-mass X-ray binary (LMXB) containing a neutron star (NS) is an old (typical age $\sim10^9\ \rm years$)  binary stellar system in which a weakly magnetised ($B\lesssim 10^{8-9}$\,G) NS accretes matter from a low-mass companion star ($\lesssim 1 M_{\sun}$) via the Roche-lobe overflow. This accretion process can be persistent or episodic. In the latter case, systems (`transients') spend most of their lives in a faint quiescent state with X-ray luminosities below $L_X\sim10^{34}\ \rm erg\ s^{-1}$. Such dormant phases are interrupted occasionally by short outbursts (typically lasting for weeks to months) in which the X-ray luminosity increases by several orders of magnitude ($L_X\sim10^{35-38}\ \rm erg\ s^{-1}$), likely due to a thermo-viscous instability in the accretion disc \citep{lasota}. On the contrary, the persistent systems are always active, with $L_X>10^{36}\ \rm erg\ cm^2\ s^{-1}$.

Based on the shape of the tracks they trace out in the X-ray colour-colour diagram (`CCD') and hardness-intensity diagram (`HID'), NS-LMXBs are often  classified into two categories: ``Z" and ``atoll" sources \citep{vanderklis1989}. These objects differ in their spectral and timing properties \citep{vanderklis2006}, with the Z sources being typically brighter ($>0.5\ L_{\rm Edd}$, where $L_{\rm Edd}$ is the Eddington luminosity) than the atoll sources ($\sim 0.001-0.5\ L_{\rm Edd}$) \citep{done2007}. 
Studies of the transient NS-LMXB XTE J$1701-462$,  which evolved through various Z source and atoll source phases   \citep{Lin2009,Homan2010}, showed that differences between the atoll and Z sources are mainly driven by the mass accretion rate (which is  higher in the Z sources). 

The broadband energy spectra of NS-LMXBs are composed of several spectral components, two of which they have in common with black hole (BH)-LMXBs: a soft thermal component due to the multi-coloured black-body emission from an optically thick and geometrically thin accretion disc, and a hard component, believed to arise due to the inverse Compton scattering of soft seed photons by a hot optically thin electron gas (`corona') located near the inner accretion flow. However, unlike the case of a BH, a NS has a solid surface in which the accreting matter is decelerated down to the angular velocity of the NS. The difference in angular momentum between the inner part of the accretion flow and the surface of the NS leads to the formation of a boundary layer (BL) \citep{shakura1988,inogamov1999} in which the kinetic energy of the accreting material is deposited \citep{gilfanov2014}. 
In Newtonian gravity, up to half of the gravitational potential energy of the accreting material is dissipated in this layer where the same parameter is even greater when general relativity is considered \citep{sibgatullin2000}. 
Thus, the X-ray emission from the BL or the surface of a NS (hereafter, we will refer to both of them as BL for convenience) significantly contributes to the X-ray spectra of NS-LMXBs, which is absent in the case of BHs. In particular, the hard state spectra for NS-LMXBs are found to be softer than the BH-LMXBs as a result of the interaction between the X-ray emission from the BL and the corona \citep{banerjee2020}.

Depending upon the relative dominance between the soft thermal component (from the disc or/and BL) and the hard Comptonized component, the spectral states of the atoll sources can be broadly divided into three different states like the BH-LMXBs: hard (`extreme island'), intermediate (`island'), and soft (`banana') state \citep{vanderklis2006}. 
On the other hand, the spectra of Z sources are usually soft \citep{vanderklis2006,mufemo2014}. 

In addition to the soft thermal and hard Comptonized components, the X-ray spectra of NS-LMXBs sometimes exhibit reflection features that are generated as a fraction of the hard Comptonized photons get scattered by the cold accretion disc. These features include a Compton hump roughly between 20-40 keV due to Compton scattering of the hard X-ray photons by the free electrons in the top layers of the disc, and few discrete features due to the fluorescence emission and photoelectric absorption by heavy ions in the disc. Out of these discrete features, the broad iron K$\alpha$ emission line around $6.4-6.97$ keV is most often observed in NS-LMXBs \citep[e.g.,][]{sudip2007,pandel2008,titarchuk2009,cackett2008,cackett2009,cackett2010,egron2013,disalvo2015,iaria2016,ludlam2017a,ludlam2017b,ludlam2019} due to a relatively high abundance of iron in the disc, with the highest fluorescence yield among the other elements. This line, which originates in the inner part of the accretion disc, is shaped by the Doppler effect from high orbital velocity and gravitational redshift due to the proximity to the compact object \citep{fabian1989,fabian2000}. As a consequence, the emission line becomes broadened and asymmetric, and is a great tool for probing the strong gravity regime. Since the relativistic effects and the Doppler boosting are stronger closer to the compact object, the iron K$\alpha$ line profile (red wing of the profile) provides a way to estimate the extent of the inner disc \citep{fabian1989,fabian2000}. This value can be used to set an upper limit for the radius of NSs as the disc must truncate before the NS surface \citep{cackett2008,cackett2010,chiang2016,degenaar2015}. It is also employed to place an upper limit on the magnetic field strength as the disc can get truncated due to the magnetic field of NSs \citep{cackett2009,chiang2016,ludlam2019}.  

The X-ray source 4U 1702$-$429 is a persistent NS-LMXB exhibiting thermonuclear X-ray bursts \citep{iaria2016}. The source was detected as a burster with \textit{OSO 8} \citep{swank1976} and  was classified as an atoll source using \textit{EXOSAT} data \citep{oosterbroek1991}. The distance to the source was estimated to be $4.19\pm0.15$ kpc or $5.46\pm0.19$ kpc for a pure hydrogen and pure helium companion star, respectively, from a photospheric radius expansion type thermonuclear burst \citep{galloway2009}. Burst oscillations were also detected at a frequency 330 Hz with \textit{RXTE} \citep{markwardt1999}, which could be considered to be identical to the spin frequency of the source \citep{deepto2003}. 

A broad iron emission line has been observed in the X-ray spectra of this source with several instruments: \textit{XMM-Newton} \citep{iaria2016,mazzola2019}, \textit{BeppoSAX} \citep{mazzola2019}, \textit{NuSTAR} \citep{ludlam2019}. \cite{iaria2016} performed the first broad-band spectral analysis of the source with \textit{XMM-Newton}/\textit{INTEGRAL} data in the energy range $0.3-60.0$ keV, and found the presence of Fe K$\alpha$ line, originating at a distance of $64_{-15}^{+52}\rm$ $R_g$ (where $R_g=GM/c^2$: $G$ is the Newton's gravitational constant, $M$ is the mass of the NS, and $c$ is the speed of the light in vacuum). Additionally, absorption edges around $0.87$ keV (related to O \texttt{VIII} ions) and $8.83$ keV (associated with Fe \texttt{XXVI} ions), were detected. Later, \cite{mazzola2019} re-analyzed the above data along with three \textit{BeppoSAX} observations and  detected three additional emission lines due to the fluorescence emission from Ar \texttt{XVIII}, Ca \texttt{XIX}, and Fe \texttt{XXV} 
in the \textit{XMM-Newton} spectrum. They constrained the inclination of the disc to be $38_{-5}^{+7}$ degree with the \textit{XMM-Newton}/\textit{INTEGRAL} data and proposed that the source was in the soft state during the \textit{XMM-Newton}/\textit{INTEGRAL} observation and in the hard state during the \textit{BeppoSAX} observations. However, in their analysis, the inner radius of the accretion disc in the hard state comes out to be smaller than the soft state, while the inner disc temperature is reported to be higher in the hard state, completely opposite to what we observe for other BH or NS-LMXBs \citep{done2007,padilla2017}. \cite{ludlam2019} also performed a reflection analysis of this source with \textit{NuSTAR} data, and found both the iron abundance ($A_{\rm Fe}$) and the ionization parameter ($\log \xi$, where $\xi=4\pi F_{\rm x}/n_{\rm e}$: $n_{\rm e}$ is the disc density and $F_{\rm x}$ is the net ionising flux) to be unusually high,  $4.9^{+4.6}_{-0.3}A_{\rm Fe,solar}$ and $4.02_{-0.03}^{+0.33}$, respectively ($A_{\rm Fe}$ was kept fixed at $2A_{\rm Fe,solar}$ in other two mentioned works).    

In this work, we perform a comprehensive spectral analysis of the atoll NS-LMXB 4U 1702$-$429 using 14 carefully selected  \textit{NICER} observations (total exposure time of $\sim 50.5$ ks). These observations roughly cover all spectral states present in the entire \textit{NICER} dataset. We probe the accretion geometry of this system in different states, by studying the evolution of the spectral parameters related to the thermal emission components (disc/BL components). Thereafter, we perform broad-band analysis of an \textit{AstroSat} observation and a \textit{NuSTAR} observation to constrain the components of reflection spectrum.
The paper is organized as follows. We describe the observations and data reduction in Section \ref{observation}. The results from the spectral analyses are presented in Section \ref{analysis}. We discuss and summarize our results in Section \ref{discuss}. 

\section{Observations and data reduction}\label{observation}
\begin{table*}
\caption{Details of the \textit{NICER} observations used in this work. Each observation is marked with an identification code based on their position in the HID/SID.}\label{tab:obs1}
  \begin{tabular}{|c|c|c|c|c|c|}
    \hline
     Obs ID & Exposure (s) $^a$ & Start Date & Position in the HID/SID\\
           &        &     yyyy-mm-dd & (ID)  \\
    \hline
         1050110113 & 588 & 2018-03-17 & 1\\
         1050110131 & 3273 & 2018-10-04  & 2\\
         1050110129 & 2230 & 2018-10-01 & 3\\
         1050110130 & 1451 & 2018-10-02  & 4\\
         1050110128 & 947 & 2018-09-30  & 5\\
         1050110118 & 742 & 2018-07-09  & 6\\
         1050110126 & 687 & 2018-09-27  & 7\\
         1050110117 & 1673 & 2018-06-13  & 8\\
         2587030103 & 15380 & 2019-08-09  & 9\\
         2587030101 & 9994 & 2019-08-08  & 10\\
         2587030104 & 9043 & 2019-08-11  & 11\\
         0050110105 & 1549 & 2017-06-28  & 12\\
         0050110107 & 1507 & 2017-06-30  & 13\\
         0050110110 & 2069 & 2017-07-07 & 14\\
    \hline
    \textbf{Note.}$^a$ Final exposure time. \\
 \end{tabular}
 \end{table*}
\begin{table*}
\caption{Details of the \textit{AstroSat} and \textit{NuSTAR} observations used in this work.}\label{tab:obs2}
  \begin{tabular}{|c|c|c|c|c|c|}
    \hline
     Instrument & Obs ID & Exposure (s) $^a$ & Start Date & Net Count Rate\\
                &        &                   & yyyy-mm-dd  & Counts $\rm s^{-1}$\\
    \hline
     \textit{AstroSat}/SXT & 20180427\_A04\_225T01\_9000002062 & 31799 & 2018-04-27 & 5\\
     \textit{AstroSat}/LAXPC &            & 8175 & & 119\\
     \hline
     \textit{NuSTAR}/FPMA & 30363005002 & 21698 & 2017-08-29 & 11\\
     \textit{NuSTAR}/FPMB &  & 21641 &  & 10\\
    \hline
    \textbf{Note.}$^a$ Final exposure time.\\
 \end{tabular}
 \end{table*}
\begin{figure}
\begin{center}
\includegraphics[width=0.45\textwidth]{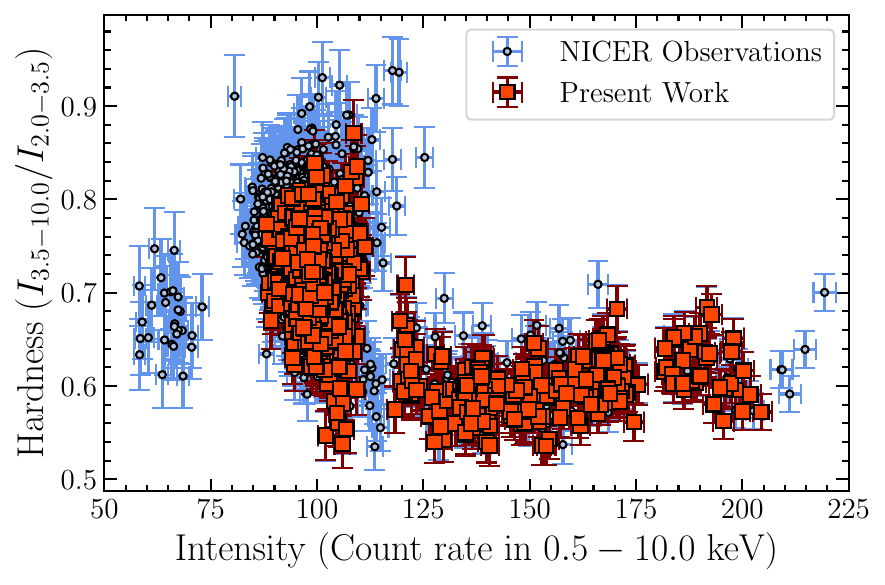}
\includegraphics[width=0.45\textwidth]{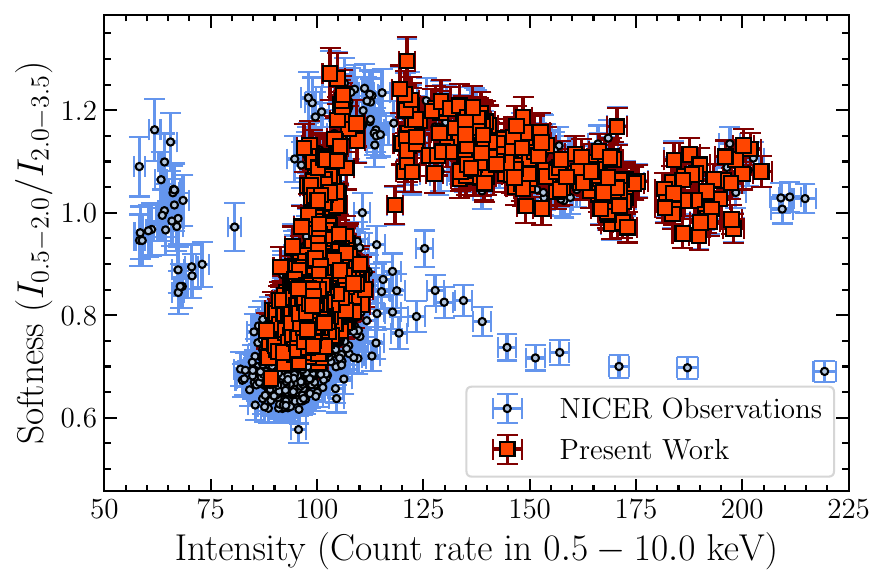}
\caption{Hardness-intensity (upper panel) and softness-intensity (lower panel) diagrams for all the public \textit{NICER} observations of the atoll NS-LMXB 4U 1702-429 with an exposure of $>100$ s, including the 14 observations considered in this work. The softness ratio is defined as the the ratio between the \textit{NICER} counts in the 2.0-3.5  keV ($I_{2.0-3.5}$) and  0.5-2.0 keV ($I_{0.5-2.0}$) bands, where hardness ratio is the ratio between the \textit{NICER} counts in the 2.0-3.5 keV ($I_{2.0-3.5}$) and 3.5-10.0 keV ($I_{3.5-10.0}$) bands. See Section \ref{observation} for more details.\label{fig:hard}}
\end{center}
\end{figure}
The data sets considered in this work are comprised of 14 \textit{NICER} observations, one \textit{NuSTAR} observation, and one \textit{AstroSat} observation.
All these observations were performed during the period of 2017-2020. The details of the observations used in this study are provided in Tables~\ref{tab:obs1} and \ref{tab:obs2}. In this work, we use the \texttt{HEASOFT} version 6.30.1 for data processing and spectral analysis.
\subsection{\textit{NICER}}
The X-ray Timing Instrument of the \textit{NICER} (Neutron Star Interior Composition Explorer) \citep{gendreau2016}, onboard the International Space Station, consists of an array of 56 co-aligned X-ray concentrator optics, each of which is paired with a single-pixel silicon drift detector that collect X-ray photons in the $0.2-12$ keV energy band with a resolution of $\sim85$ eV at 1 keV and $\sim137$ eV at 6 keV. Although 52 detectors are working presently, we do not consider data from the detectors numbered 14 and 34 as they often show periods of increased noise. In this work,  \textit{NICER} data are reduced and calibrated using the \texttt{NICERDAS} 2022-01-17\_V009 and \texttt{CALDB} version xti20210707. 

We first download all the \textit{NICER} observations available for this source with an exposure time of $>100$s, and generate cleaned event files using the standard tool \texttt{nicerl2}. We produce background uncorrected \textit{NICER} light-curves in four energy bands: $0.5-2.0$ keV, $2.0-3.5$ keV, $3.5-10.0$ keV, and $0.5-10.0$ keV (with 32s bin time) to obtain the hardness-intensity and softness-intensity (see Fig. \ref{fig:hard}) diagrams. In our work, the softness ratio is defined as the ratio between the \textit{NICER} counts in the 2.0-3.5  keV and  0.5-2.0 keV bands, where hardness ratio is the ratio between the \textit{NICER} counts in the 2.0-3.5 keV and 3.5-10.0 keV bands. From the entire dataset, we select 14 observations (each of them has an exposure time of $>500$s) such that it faithfully covers all spectral states present in the dataset (see Table~\ref{tab:obs1} and Fig.~\ref{fig:hard}). Since there is significant variation in intensity/colour during the observation period of 1050110128, 1050110129, 1050110130, we remove some of the time intervals from these observations that overlap with other selected observations in the HID.
We also obtain the mean hardness - mean intensity diagram (HID) and mean softness-mean intensity diagram (SID), and designate these observations by numbers, according to their position in the HID/SID (see Fig.~\ref{fig:mhard} and Table.~\ref{tab:obs1}). 
We refer to those observations as soft state (or banana) observations, when the softness $>1.0$ and the hardness $<0.7$ (observations 1-8). These include lower and upper banana (hereafter, LB and UB, respectively) state observations (see Fig.~\ref{fig:mhard} and Table.~\ref{tab:obs1}). The other observations (observations 9-14) are likely  in the intermediate state (perhaps extending into the   hard state, which is difficult to ascertain due to a relatively low upper pass-band ($10$ keV) of the \textit{NICER} data.

We employ the background estimator \texttt{nibackgen3C50} \citep{remillard2021} to generate the source and background spectra, and use \texttt{nicerarf} and \texttt{nicerrmf} to obtain the arf and rmf files for these 14 \textit{NICER} observations. We do not detect a thermonuclear X-ray burst in any of these observations. We group the \textit{NICER} spectra with an exposure time of $<1$ ks and $>1$ ks using the FTOOLS package \texttt{ftgrouppha} to a signal-to-noise ratio (SNR) of at least 15 per bin and 20 per bin, respectively.

\subsection{\textit{NuSTAR}}
The details of \textit{NuSTAR} observation are provided in Table~\ref{tab:obs2}.
We follow the standard data reduction procedure using \texttt{NUSTARDAS} v2.1.2 with \texttt{CALDB} 20220926 for processing the \textit{NuSTAR} data. We generate the source light-curve and spectrum from a circular region with a radius of 120'' centered around the source. For background, we consider a 80'' radial region far away from source. No thermonuclear burst is detected during this observation. We group the \textit{NuSTAR} spectrum with a SNR of minimum 15 per bin.

\subsection{\textit{AstroSat}}
\textit{AstroSat} \citep{singh2014}, the first Indian dedicated astronomy satellite, offers an unique opportunity to observe X-ray binaries in a broad-band energy range of 0.5–80.0 keV with its two co-aligned instruments: Soft X-ray Telescope (SXT) and Large Area X-ray Proportional Counter (LAXPC). SXT \citep{singh2016orbit,singh2017soft} is a focusing X-ray telescope with a CCD camera and operates in the photon counting mode. It is suitable for performing a medium resolution X-ray spectroscopy (full width at half maximum, FWHM, $\sim150$ eV at 6 keV) in the energy range 0.5-7.0 keV and also provides low resolution imaging (FWHM $\sim2'$).
We process the level-1 data using the SXT pipeline AS1SXTLevel2-1.4b\footnote{\url{https://www.tifr.res.in/~astrosat_sxt/sxtpipeline.html}} and obtain level-2 clean event files for individual orbits. We thus merge the orbit-wise clean event files into a single event file with a net exposure of $\sim 32$ ks using the Julia-based SXT event merger tool SXTMerger.jl\footnote{\url{https://github.com/gulabd/SXTMerger.jl}}. Using the \texttt{XSELECT} tool, we obtain the source light-curve from a circular region of 15'' centred on the centre, and detect a thermonuclear X-ray burst. Thus, we determine the time interval at which the burst occurred, exclude it, and obtain a burst-excluded spectrum with an exposure of $\sim 31.8$ ks from the same region (see Table~\ref{tab:obs2}). We use the standard deep blank sky background spectrum (SkyBkg\_comb\_EL3p5\_Cl\_Rd16p0\_v01.pha), and the response matrix file (RMF) (sxt\_pc\_mat\_g0to12.rmf), provided by the instrument team for spectral analysis with the SXT data \footnote{\url{http://astrosat-ssc.iucaa.in/sxtData}}. In addition, we utilize an updated ancillary response file (ARF) \citep{pasham2022,swain2023} in this work.

The X-ray instrument LAXPC \citep{yadav2016,antia2017} comprises of three proportional counters (LAXPC10, LAXPC20, and LAXPC30) with an effective area of 6000 $\rm cm^2$ at 15 keV and has a time resolution of 10$\mu$s in the energy range 3.0-80.0 keV. Out of these three detectors, LAXPC10 has been showing abnormal gain variation from 2018, March and LAXPC30 was shut off due to a gas leakage \citep{antia2021}. Thus, we only use the data acquired with the detector LAXPC20 in this work. We extract the source and background light curves and spectrum using the software \texttt{laxpcsoftv3.4.3} \footnote{\url{https://www.tifr.res.in/~astrosat_laxpc/LaxpcSoft_v1.0/antia/laxpcsoftv3.4.3_07May2022.tar.gz}} and use the response file, lx20v1.0.rmf, for spectral analysis. We do not see any thermonuclear bursts during the LAXPC observation period.
We group the SXT and LAXPC spectra using the FTOOLS package \texttt{ftgrouppha} to a SNR of at least 12 per bin and 15 per bin, respectively.
We add a systematic uncertainty of $2\%$ to each bin of the SXT and LAXPC spectra, as suggested by their respective instrument teams. 
\section{Spectral Analysis and Results}\label{analysis}
We use \texttt{XSPEC} version 12.12.1 \citep{arnaud1996} to perform spectral analyses in this work. The energy ranges of $0.3-10.0$ keV, $1.3-6.0$ keV, $4.0-25.0$ keV, and $3.0-50.0$ keV are considered for \textit{NICER}, SXT, LAXPC, and \textit{NuSTAR} data, respectively. 
For performing joint spectral fitting of the SXT and LAXPC data, a cross-normalization constant (\texttt{constant} in \texttt{XSPEC}) is allowed to vary freely for LAXPC and fixed to unity for SXT. Similarly, we use a cross-normalization constant, which is kept fixed to unity for FPMA and vary freely for FPMB, while performing analyses with the NuSTAR data. We apply a gain correction to the SXT data using the \texttt{XSPEC} command \texttt{gain fit} with slope fixed to unity. The best-fit offset is found to be $\sim 0.3$ eV.
We use \texttt{tbabs} \citep{wilms2000} to account for the photoelectric absorption by the neutral matter in the ISM along the line of sight. We set the abundances to  \texttt{wilm} \citep{wilms2000} and the photoelectric absorption cross-sections to \texttt{vern} \citep{vern1996}. Furthermore, we assume a distance to the source of 5.5 kpc \citep{iaria2016} and a NS mass of $1.4 M_{\sun}$ in our analysis.  All parameter uncertainties reported here are at the $90\%$ confidence level for one parameter of interest.
\begin{figure}
\includegraphics[width=0.45\textwidth]{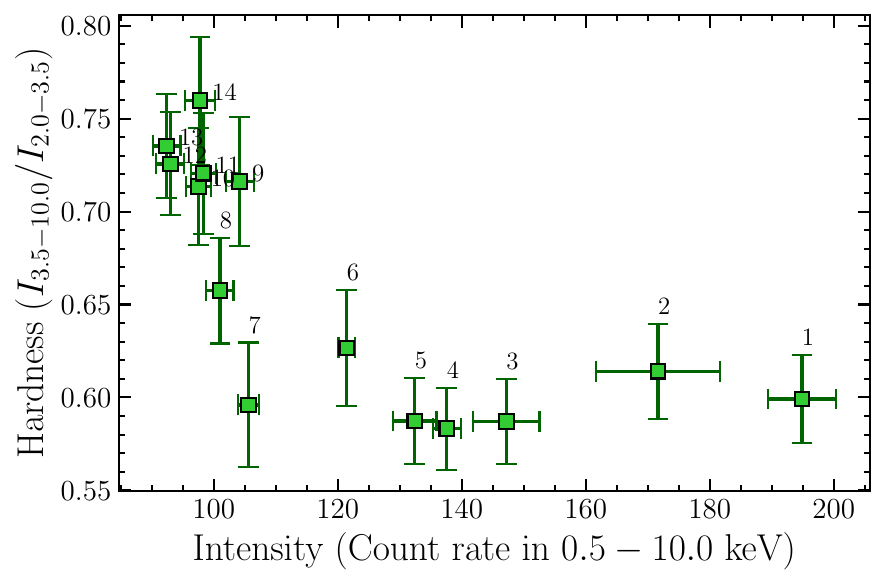}
\includegraphics[width=0.45\textwidth]{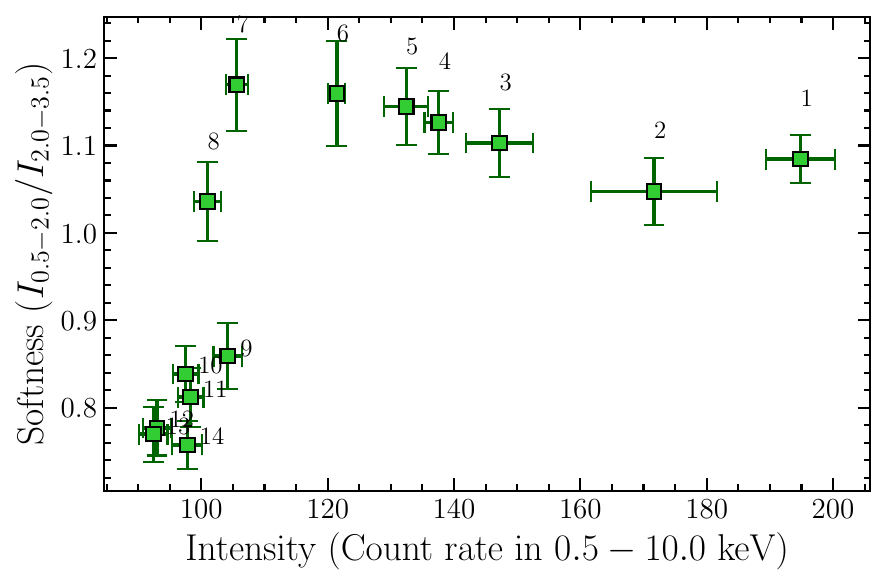}
\caption{Hardness (mean)-Intensity (mean)  diagram (HID) and Softness (mean) - Intensity (mean) diagram (SID) for the \textit{NICER} observations of the atoll NS-LMXB 4U 1702-429 used in this work. The hardness ratio is the ratio between the \textit{NICER} counts in the 2.0-3.5 keV ($I_{2.0-3.5}$) and 3.5-10.0 keV ($I_{3.5-10.0}$) bands, where the softness ratio is the ratio between the \textit{NICER} counts in the 2.0-3.5 keV ($I_{2.0-3.5}$) and 0.5-2.0 keV ($I_{0.5-2.0}$) bands. The observations are marked by numbers, according to their position in HID/SID. See Section \ref{observation} for more details.\label{fig:mhard}}
\end{figure}
\subsection{Analysis of the \textit{NICER} spectra}\label{sec:nicer}
We begin our investigation with the \textit{NICER} soft/banana state spectra (observations 1-8) in the energy band 0.3-10.0 keV with a three component model. Our model consists of a multicoloured disk blackbody \texttt{diskbb} \citep{mitsuda1984,makhishima1986} (for representing the soft disk emission), a single temperature black-body (BB) component \texttt{bbody} (for describing the X-ray emission from the BL), a thermal Comptonization component (accounting for the hard coronal emission) \texttt{nthcomp} \citep{zdziarski1996} with the parameter \texttt{inp\_type} of \texttt{nthcomp} being set to 0 (indicating that the soft seed photons are supplied by the BL). Thus, we tie the seed photon temperature $kT_{\rm seed}$ of the \texttt{nthomp} component with the \texttt{bbody} temperature $kT_{\rm BB}$. So, our proposed three component model has 8 independent spectral parameters: neutral hydrogen column density $N_{\rm H}$, disc temperature $kT_{\rm in}$, disc normalization $N_{\rm diskbb}$, BB temperature $kT_{\rm BB}$, BB normalization $N_{\rm bbody}$, power-law index $\Gamma$, electron temperature $kT_{\rm e}$, and \texttt{nthcomp} normalization, which are to be constrained from the spectral fit. \cite{lin2007} considered a similar model for the banana state observations in their work, where they used a constrained broken power-law for the hard component. In our work, we opt a more physically meaningful \texttt{nthcomp} component for representing the Comptonized X-ray emission. Interestingly, our adopted model was earlier used to describe the 0.8-30.0 keV \textit{Suzaku} spectra of the atoll NS 4U 1608-429 \citep{padilla2017}.

Due to the absence of hard X-ray data (> 10.0 keV) in our spectral analyses, it becomes difficult to find a consistent model to describe the weak Comptonized hard emission in the banana state. We find that the addition of the \texttt{nthcomp} component is not statistically significant for observations 1 and 2 (i.e., the UB observations). 
After experimenting with several Comptonized components, we resort to using a pegged power-law (\texttt{pegpwrlw}) to represent the hard component of observation 2, but we could not find a suitable model for observation 1, the inclusion of which would be significant. So, we finally ended up with only a two-component model for observation 1 (i.e., without a Comptonization component).
Besides, both the electron temperature $kT_{\rm e}$ and the power-law index $\Gamma$ remain unconstrained in their reasonable limits for observations 3-7 (we restrict the upper limit of $kT_{\rm e}$ and $\Gamma$ to 100 keV and 3, respectively). So, we fix the electron temperature $kT_{\rm e}$ to 10 keV and $\Gamma$ to 2 for these observations. We also try other Comptonized components like \texttt{thcomp} \citep{zdziarski2020} or \texttt{simpl} \citep{steiner2009} for  observations 2-8, but the spectral fits always give unphysically large values ($>0.5$) of Comptonized fraction. The problems related to the modelling of the hard component are  due to \textit{NICER}'s lack of high-energy coverage. Therefore, we use the following models to fit the 8 banana state observations:
\begin{itemize}
    \item Model A: \texttt{tbabs}*(\texttt{diskbb}+\texttt{bbody}) [for obs. 1]
    \item Model $\mathscr{A}$: \texttt{tbabs}*(\texttt{diskbb}+\texttt{bbody}+\texttt{pegpwrlw})  [for obs. 2]
    \item Model B: \texttt{tbabs}*(\texttt{diskbb}+\texttt{bbody}+\texttt{nthcomp[inp\_type=0]}) [for obs. 3-8]
\end{itemize}
Due to the large residuals around 0.85 keV for  observations 2, 3, 4, 5, and 8, we add an absorption edge (\texttt{edge} in \texttt{XSPEC} notation) to our previously mentioned models. This feature is most likely associated with \ion{O VIII}, and such a component was also employed earlier for analysing the \textit{XMM-Newton} spectra of 4U 1702-429 \citep{iaria2016,mazzola2019}. In all the cases, the neutral hydrogen column density ($N_{\rm H}$) lies in the range of $2.1-2.3\times10^{+22}\ \rm atoms\ cm^{-2}$, which is close to the earlier estimated value of $\sim 2.3-2.5\times10^{+22}\ \rm atoms\ cm^{-2}$ \citep{iaria2016,mazzola2019}. Hereafter, we fix this quantity to the median value $2.2\times10^{+22}\ \rm atoms\ cm^{-2}$ for all the spectral analyses performed in this work. We do not detect reflection features in any of the banana state observations with $\geq 3\sigma$ significance.

We thereafter consider the island/intermediate state observations (obs. 9-14) and fit the spectra with the same three component model mentioned above. However, the addition of the \texttt{bbody} component becomes statistically insignificant for these observations. Additionally, we detect iron K-$\alpha$ emission line in  observations 10 and 11, with a significance of $\sim 3-4\sigma$. Since the emission line is weak, we employ a Gaussian (\texttt{gauss} in \texttt{XSPEC} notation) to represent this emission line for these two observations. We also try to fit the Fe K-$\alpha$ line with the relativistic emission line component \texttt{relline} in  observations 10 and 11. However, this model results in a worse fit than the basic Gaussian description.
While we do not require to fix $kT_{\rm e}$ and $\Gamma$ for these observations, we need to add an absorption edge around 0.87 keV for observations 9-11 like some of the banana state observations. Therefore, our models for the island/intermediate state spectra take the following form:
\begin{itemize}
    \item Model C: \texttt{tbabs}*(\texttt{diskbb}+\texttt{nthcomp[inp\_type=0]}) [for obs. 9,12-14]
    \item Model $\mathscr{C}$: \texttt{tbabs}*(\texttt{diskbb}+\texttt{gauss}+\texttt{nthcomp[inp\_type=0]})  [for obs. 10,11]
\end{itemize}
Although our model for the island/intermediate state spectra is different from that of \cite{lin2007} (they considered a single temperature blackbody component, and a broken power-law), both \cite{iaria2016} and \cite{mazzola2019} considered this model for describing the broad-band continuum of this source in their analyses of the \textit{XMM-Newton}/INTEGRAL and \textit{BeppoSAX} data.
 
\begin{figure}
\includegraphics[width=0.45\textwidth]{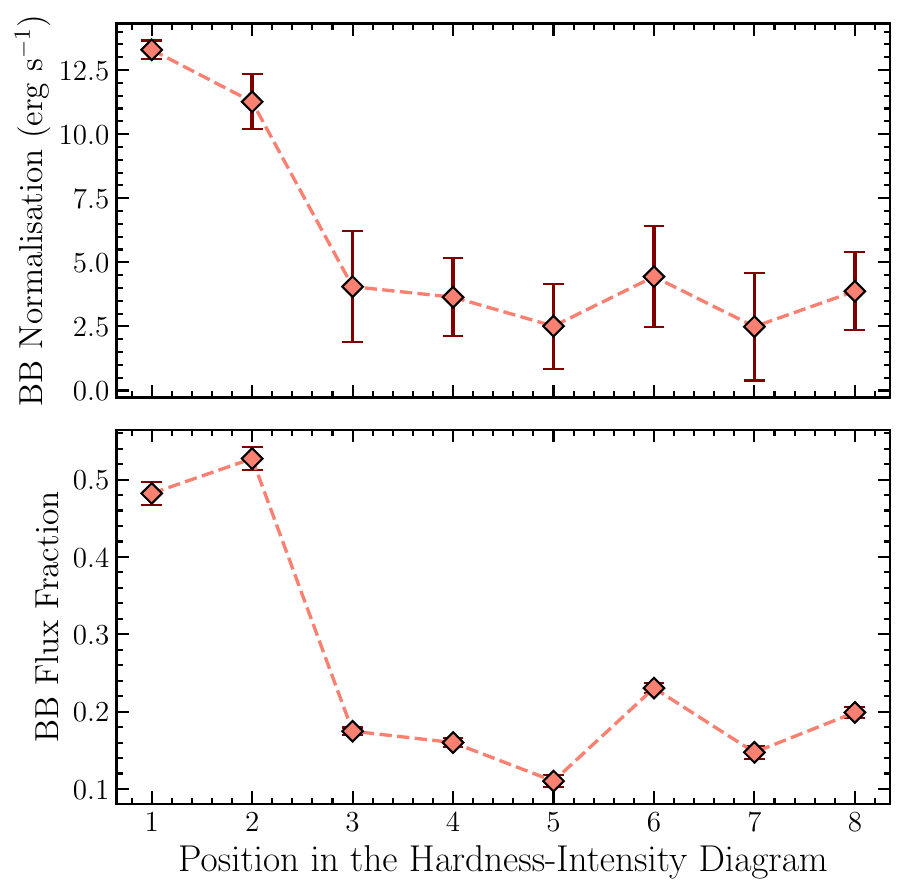}
\caption{Blackbody (BB) flux fraction (lower panel) and normalization of the blackbody component (upper panel) for the 8 soft state (banana state) \textit{NICER} observations used in this work. All fluxes are calculated in the energy range of 0.1-50.0 keV. The BB normalization is scaled in the unit of $L_{36}/D_{10}^2$, where $L_{36}$ is the luminosity in the unit of $10^{36}\ \rm erg \ s^{-1}$ and $D_{10}=D/10\rm kpc$ ($D$ is the distance to the source).
See Section \ref{sec:nicer} for more details.\label{fig:bb}}
\end{figure}

\begin{figure}
\includegraphics[width=0.45\textwidth]{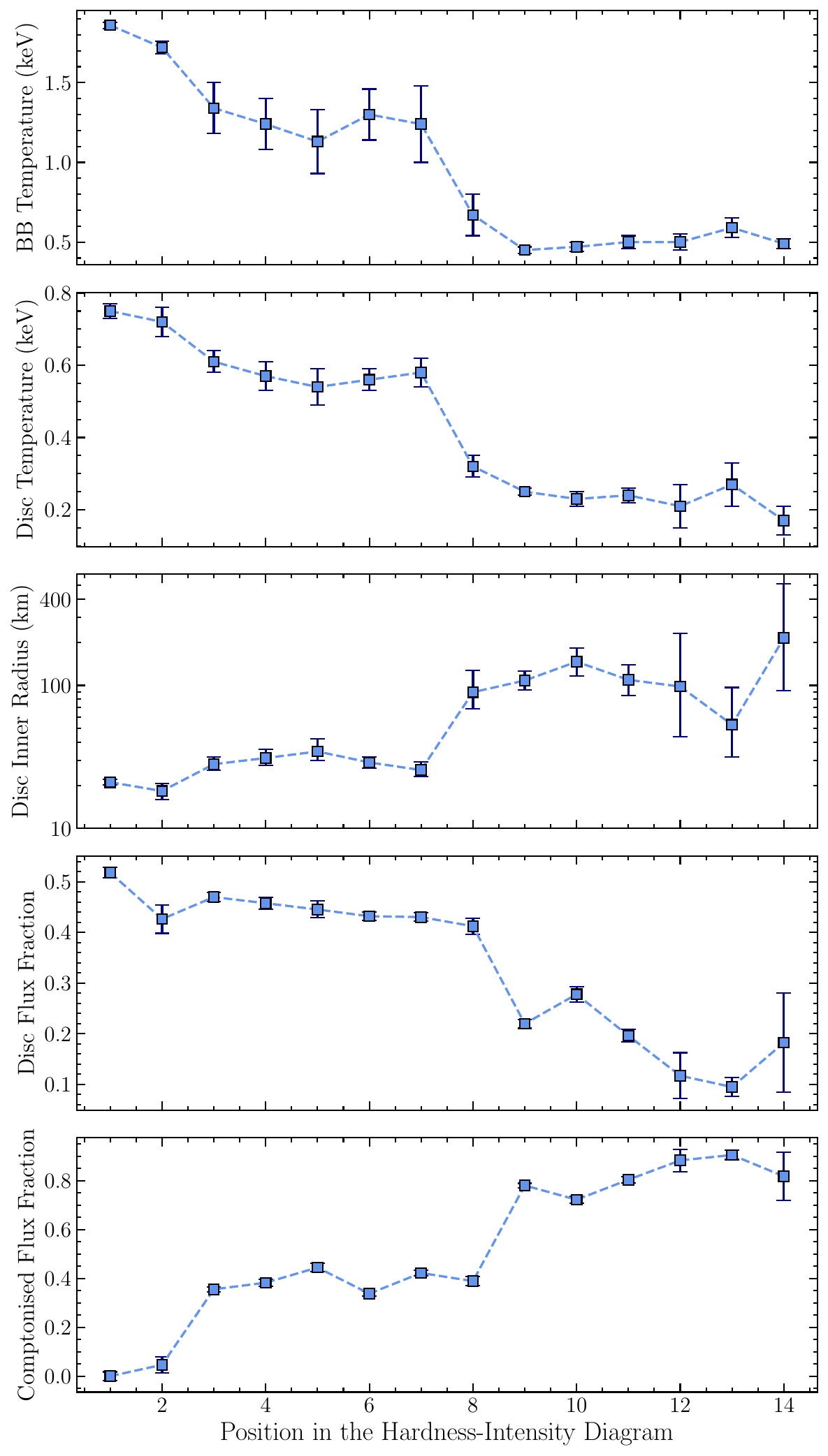}
\caption{Blackbody (BB) temperature (keV), disc temperature (keV), disc true inner radius (km), disc flux fraction, and Comptonized flux fraction for the 14 \textit{NICER} observations are depicted from top to bottom panel, respectively. All fluxes are calculated in the energy range of 0.1-50.0 keV. Here, the true inner radius of the disc is given by $r_{\rm in}=\kappa^2R_{\rm in}$, where $\kappa$ represents the colour correction factor, and $R_{\rm in}$ is the inner radius of the disc calculated from the \texttt{diskbb} normalization. This plot exhibits the evolution of several spectral parameters across different spectral states. See Section \ref{sec:nicer} for more details.\label{fig:spec}}
\end{figure}

\begin{figure*}
    \includegraphics[width=0.4\textwidth]{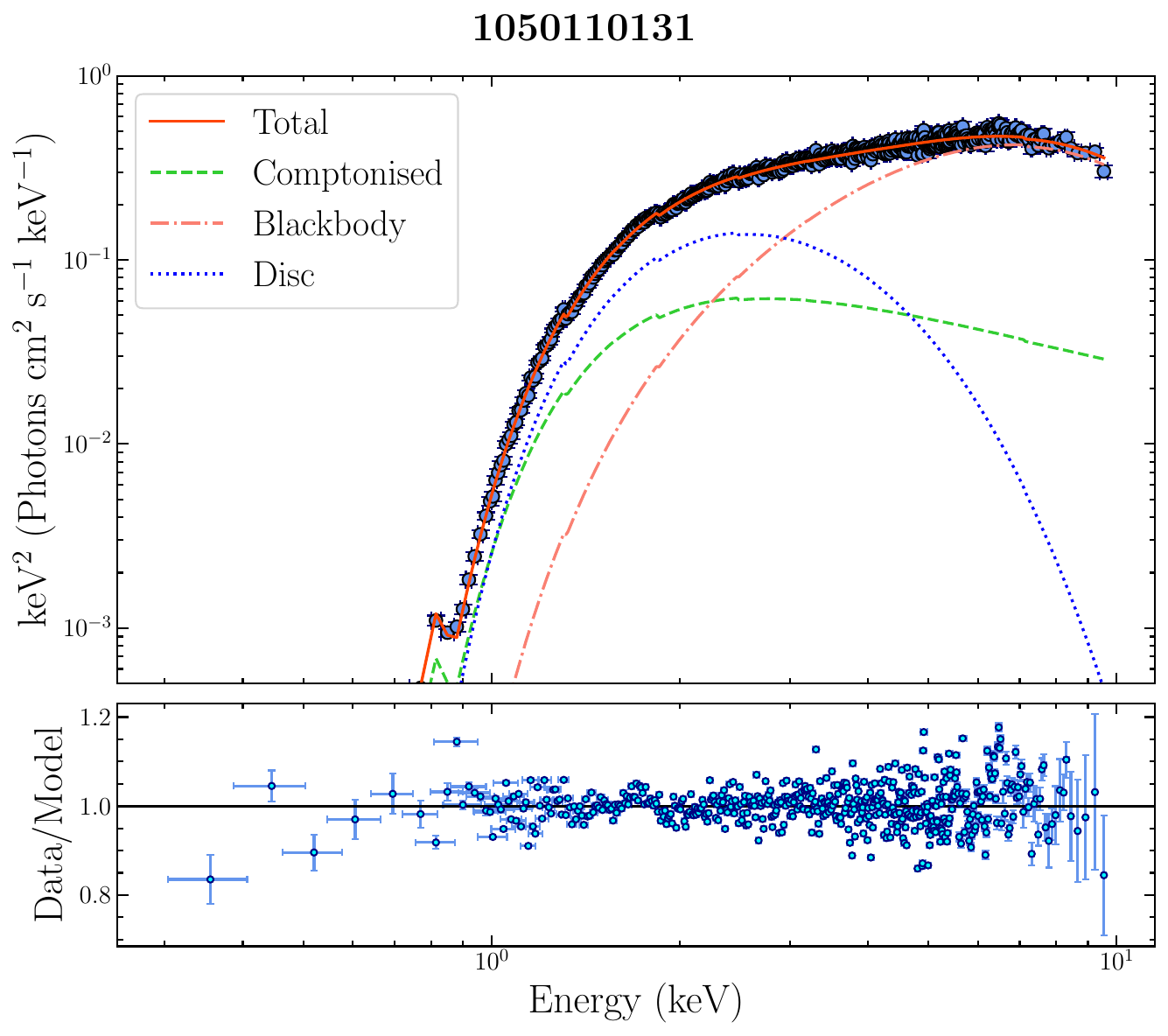}
    \includegraphics[width=0.4\textwidth]{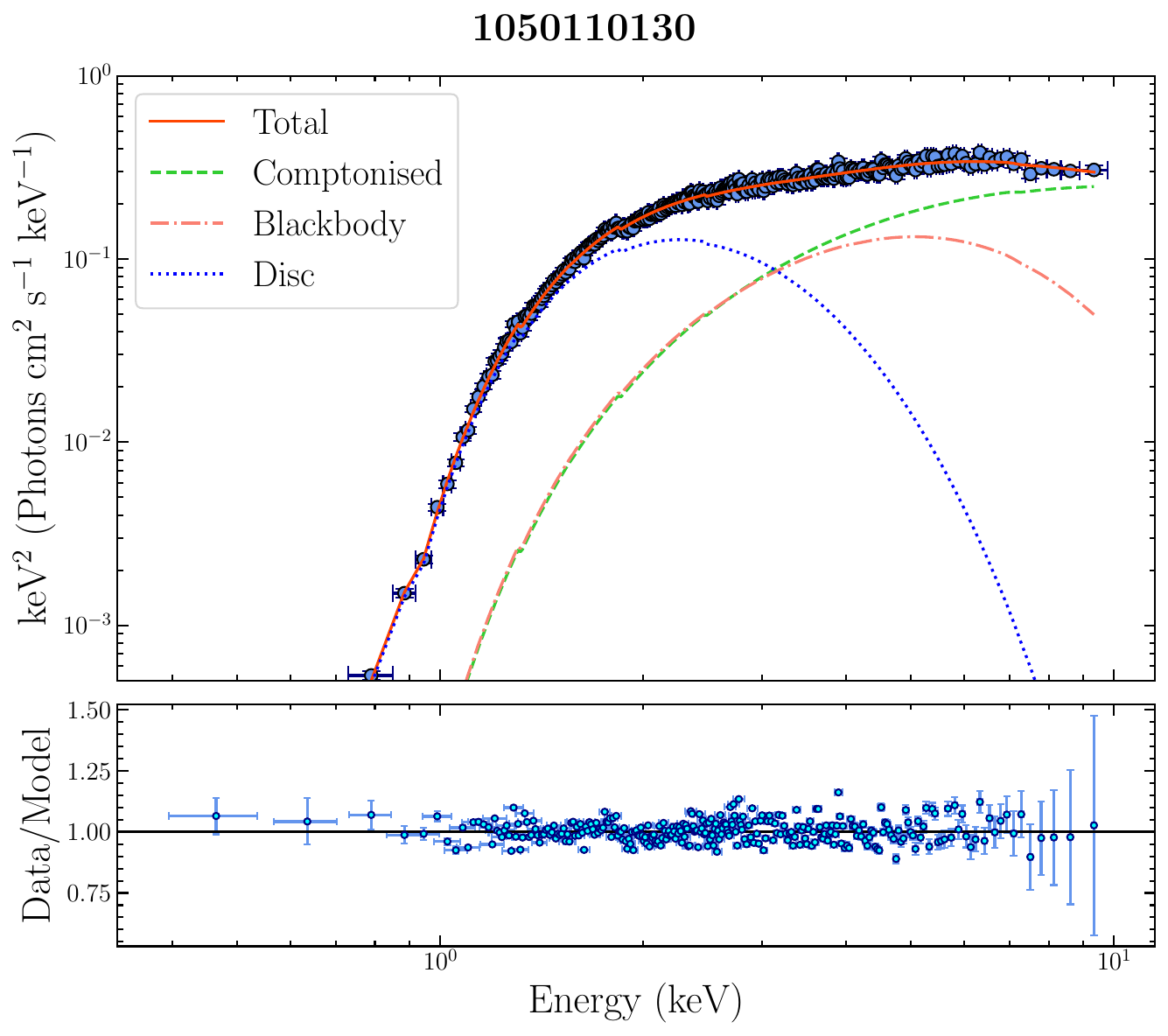}
    \includegraphics[width=0.4\textwidth]{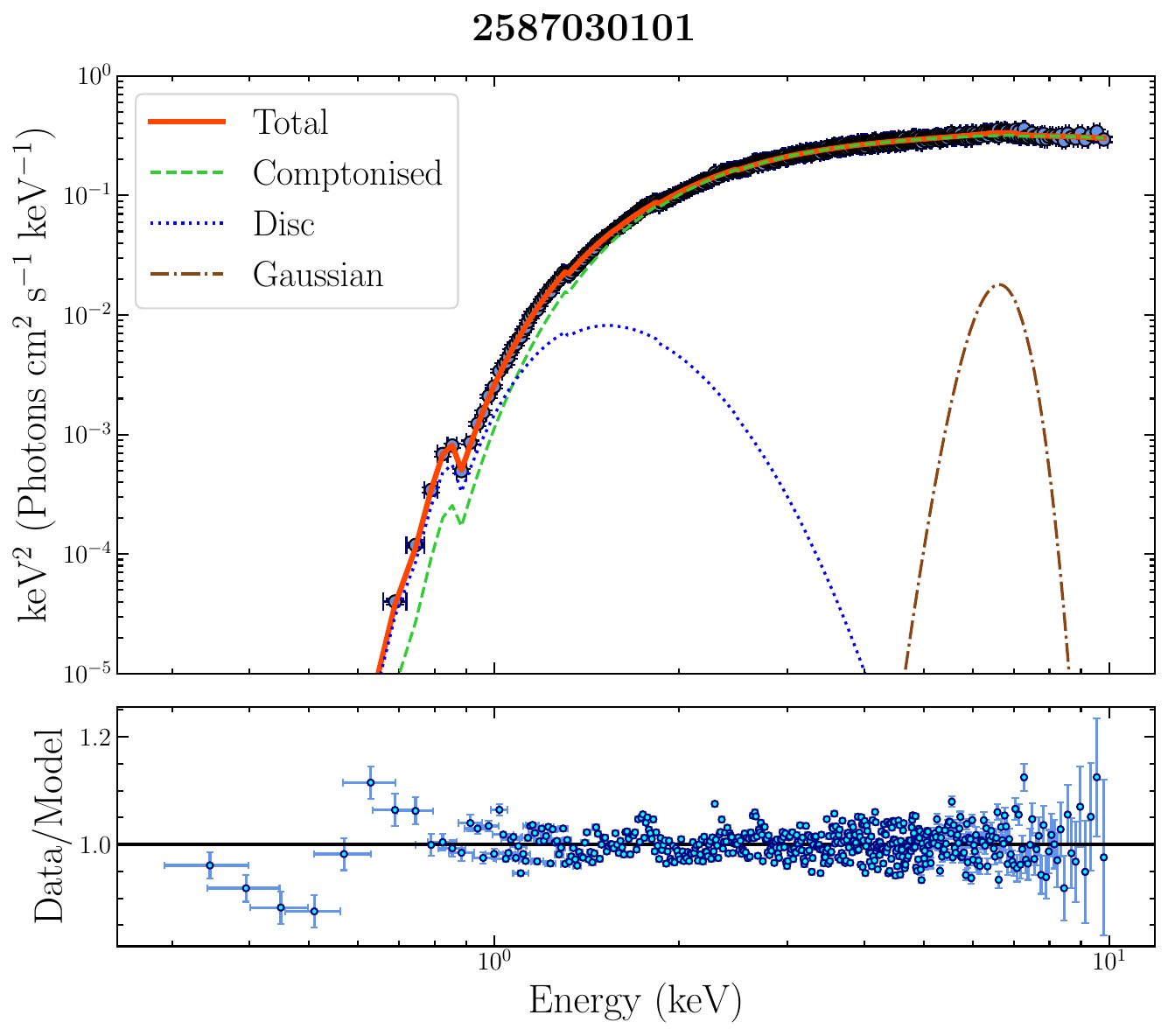}
    \caption{Unfolded spectra with different model components, and ratio of the 0.3-10.0 keV data to the best-fit model for three representative observations: 1050110131, 1050110130, and 2587030101 (observations 2, 4, and 10, respectively in Fig.~\ref{fig:mhard}). The first two observations belong to the banana state, whereas the third one falls into the island/intermediate state. Data are rebinned for plotting purpose. See Section \ref{sec:nicer} for more details.\label{fig:unfold}}
\end{figure*}

The models (with constraints mentioned above) provide a satisfactory fit to the \textit{NICER} spectra for all the observations with $\chi^2/$d.o.f (d.o.f refers to degrees of freedom) in the range of 0.9-1.2. The results are presented in Table~\ref{tab:nicer}. The Gaussian component of observation 10 has an energy of $6.55\pm0.19$ keV, a width of $0.51\pm0.24$ keV, and a normalization of $5.48\pm2.95\times10^{-4}$, whereas the same quantities for observation 11 have the values $6.58\pm0.33$ keV, $0.78\pm0.44$ keV, and $6.94\pm8.21\times10^{-4}$, respectively. The corresponding equivalent widths for these two observations are $72_{-32}^{+34}$ eV and $89_{-70}^{+26}$ eV, respectively.
We also compute the unabsorbed BB flux, disc flux, and the total flux in the energy range of 0.1-50.0 keV, by convolving \texttt{cflux} over the respective spectral component(s). Here, we note that the flux values of the disc and BB components contribute negligibly beyond 10 keV and 20 keV, respectively.
The flux values for all these observations are provided in Table~\ref{tab:nicerflux}. Since the power-law flux diverges at low energies \citep{steiner2009}, computing the total flux in the energy band 0.1-50.0 keV in this manner becomes problematic for observation 2. Therefore, we restrict the \texttt{pegpwrlw} flux to contribute only in the energy band 4.0-50.0 keV (we choose these values as the lower-peg and higher-peg values, respectively) for this observation. The evolution of several spectral parameters across the spectral states is depicted in Figures~\ref{fig:bb} and \ref{fig:spec}. The unfolded spectra (with different spectral components) of three representative observations, belonging to UB, LB, and island/intermediate states, are shown in Figure~\ref{fig:unfold}.

To check the consistency of our result, we estimate the true inner radius of the disc ($r_{\rm in}$) from the \texttt{diskbb} normalization, $N_{\rm diskbb}$. The disc inner radius, $r_{\rm in}$, must have a comparable or higher value than the size of the NS \citep{lin2007}.
We first convert $N_{\rm diskbb}$ to the disc apparent inner radius ($R_{\rm in}$) by considering the relation: $N_{\rm diskbb}=\left(R_{\rm in}/D_{10})^2\right.\cos\theta$ \citep{kubota1998}, where, $\theta$ is the disc inclination angle and $D_{10}$ is the distance to the source in the unit of 10 kpc. We take into account the colour correction factor $\kappa$ to obtain the value of $r_{\rm in}=\kappa^2R_{\rm in}$ from $R_{\rm in}$ \citep{mazzola2019}. Here, we do not consider the correction factor for the inner torque-free boundary condition because this condition is most likely not satisfied in accreting NSs \citep{frank2002}.
Assuming $D=5.5$ kpc \citep{iaria2016}, $\theta=38^{\circ}$ \citep{mazzola2019}, $\kappa=1.7$ \citep{shimura1995}, we obtain $r_{\rm in}\geq16$ km, which is higher than the typical radius ($\simeq10$ km) of a NS. 
Furthermore, our results also satisfy the criterion that the BB temperature must be higher than the disc temperature for all the observations \citep{popham2001,lin2007}.

In our three component model (i.e., Model B) or two component models (i.e, Model C and Model $\mathscr{C}$), we consider that the soft seed photons are provided by the BL. However, the seed photons for Comptonized emission can also be supplied by the disc. So, it is instructive to check how much our results change if we assume the latter scenario. For this purpose, we first set the parameter \texttt{inp\_type} of the \texttt{nthcomp} component to 1 and tie $kT_{\rm seed}$ to $kT_{\rm in}$ in our Model B (hereafter, we will call this as Model $\rm B^{*}$). We find that the Comptonized flux fraction for the LB state observations (obs. 3-8) is $>0.5$, which is improbable.
Therefore, the assumption that the soft seed photons are supplied by the BL, not the disc, is preferred due to physical consistency for the banana state observations.
If we employ Model $\rm B^{*}$ for our island/intermediate state observations (obs. 9-14) to describe the continuum, the \texttt{diskbb} component becomes statistically redundant. This new model (hereafter, we call it Model $\rm C^{*}$) gives a similar $\chi^2$/d.o.f as Model C (or Model $\mathscr{C}$). $kT_{\rm in}$ ($kT_{\rm BB}$) in this new model takes a slightly smaller (higher) value than that obtained with Model C (or Model $\mathscr{C}$), but $kT_{\rm in}$ becomes poorly constrained for observations 12-14.  Therefore, our adopted model for island/intermediate state observations provides a better description of the continuum than Model $\rm C^{*}$. 
 
In all the \textit{NICER} observations mentioned above, the discrete reflection (such as the iron K-$\alpha$ emission line) features are either absent (obs. 1-9, 12-14) or weak (obs. 10,11). Now, to constrain the parameters of the reflected spectrum consistently, we need broad-band X-ray data, i.e., data with high energy coverage beyond 10 keV. 
Thus, in order to probe the reason behind the absence (or weak presence) of the reflection feature in our \textit{NICER} sample, we consider \textit{NuSTAR} and \textit{AstroSat} broad-band observations of this source, which we analyse in the following subsections. 
\subsection{Analysis of the \textit{AstroSat} spectra}\label{sec:astrosat}
\begin{figure}
    \includegraphics[width=0.4\textwidth]{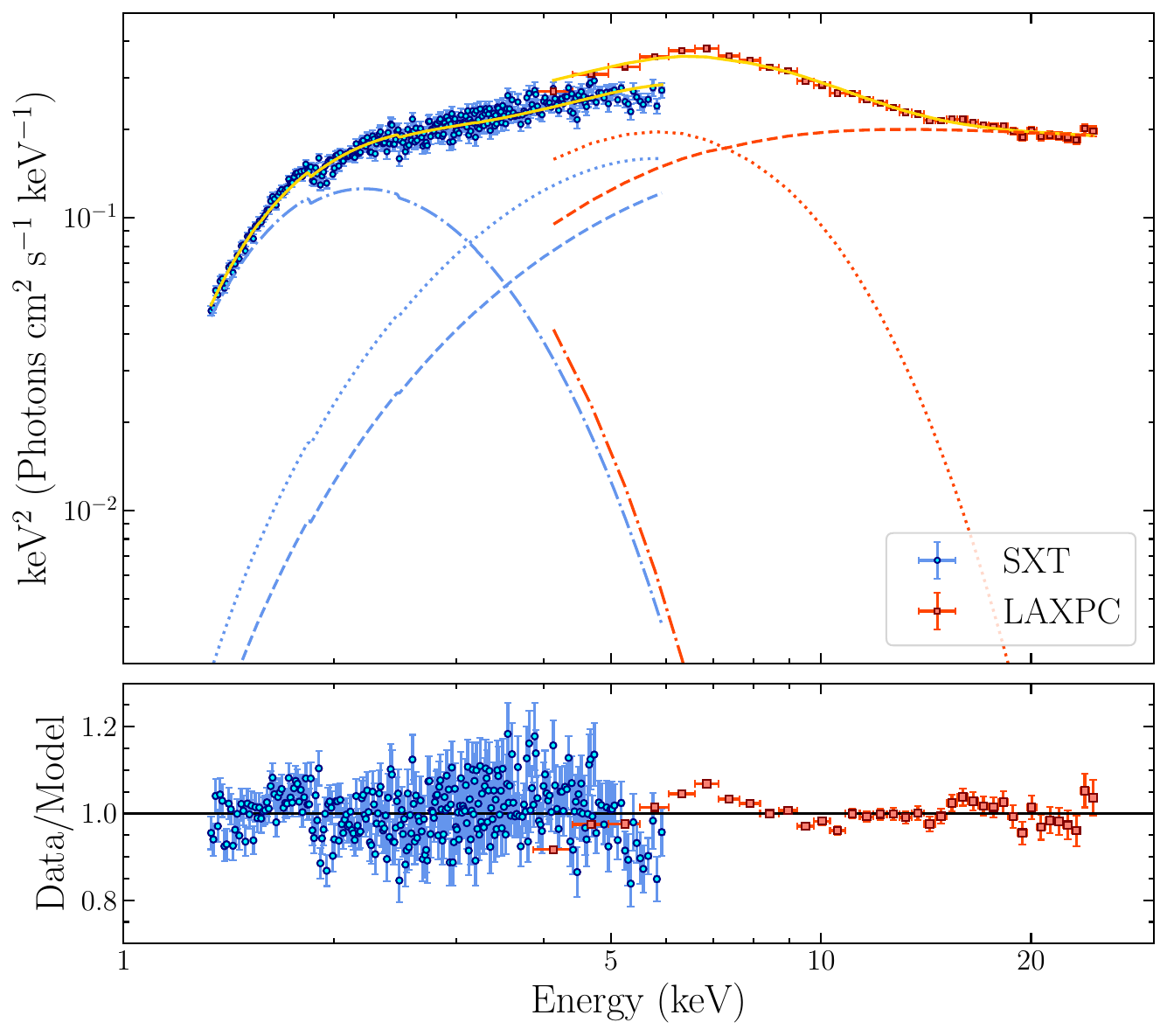}
    \caption{Unfolded spectra with different model components, and ratio of the 1.3-25.0 keV data to the model, Model B, for the \textit{AstroSat} observation. The dotted, dotted-dashed, and dashed lines refer to \texttt{bbody}, \texttt{diskbb}, and \texttt{nthcomp} components, respectively. The full unfolded spectra is represented by an orange solid line. Data are rebinned for plotting purpose. See Section \ref{sec:astrosat} for more details.\label{fig:astrounfold}}
\end{figure}

We first fit the joint SXT and LAXPC spectra in the energy band of 1.3-25.0 keV with our three component model, Model B. 
The spectral fit provides a $\chi^2/$d.o.f of 482.8/372. Here, we fix the value of $kT_{\rm e}$ to 100 keV as it could not be constrained.
The results are provided in Table~\ref{tab:astrosat}, and unfolded spectra and residuals are shown in Fig.~\ref{fig:astrounfold}. 
We find that the disc ($kT_{\rm in}=0.56\pm0.02$ keV) and BL temperature ($kT_{\rm BB}=1.44\pm0.05$ keV) of this observation are consistent with the \textit{NICER} lower-banana state (LB state) observations. The 0.1-50.0 keV unabsorbed flux of this observation is $\sim18.5\times10^{-10}\ \rm erg\ cm^{-1} s^{-1}$, and the disc and BL together provide $\sim65\%$ of the X-ray photons in the 0.1-50.0 keV band. Therefore, we can conclude that the \textit{AstroSat} observation belongs to the banana state (most likely LB state) observation, by comparing our present result with our analyses with \textit{NICER} data. In order to check the existence of discrete reflection features (based on the broad residuals present in Fig.~\ref{fig:astrounfold}), we add a Gaussian line, \texttt{gauss}, around 6.5 keV to our existing model, and find this addition to be $<3\sigma$ significant. Besides, the added Gaussian line takes almost the shape of a continuum, and affects the \texttt{bbody} components significantly, making the value of the relevant spectral parameters unreliable. Therefore, like the \textit{NICER} banana state observations, we do not detect discrete reflection features (like Fe-K$\alpha$ emission line) significantly in the \textit{AstroSat} observation.

We now employ a reflection model to investigate the reason behind the lack of reflection features in the data.
We replace the \texttt{nthcomp} component in Model B with the self-consistent reflection component \texttt{reflionxhc} from the \texttt{reflionx} suite \citep{ross2005,ross2007} (hereafter referred to as Model D), which models the reflection spectrum along with describing the incident hard continuum. The \texttt{reflionx} based reflection models produce an angle-averaged reflection spectrum for an optically thick atmosphere (i.e., the surface of an accretion disc) of constant density illuminated by the hard Comptonized emission. The component \texttt{reflionxhc} assumes that the incident continuum is described by a cutoff power-law. Here, we fix the value of high-energy cutoff ($E_{\rm c}$) to 300 keV (as $E_{\rm c}$ was pegged at the model maximum value of 600 keV) and $N_{\rm H}$ to $2.2\times10^{22}\ \rm atoms\ cm^{-2}$, i.e., our chosen value in our spectral analyses with \textit{NICER} data (which also lies in the estimated range of this parameter; see Table~\ref{tab:astrosat}). 
We obtain a significant improvement (compared to our Model B) in the spectral fit with a $\chi^2/$d.o.f of 462.9/371. The results are presented in Table~\ref{tab:astrosat}. Since the Comptonized component of Model B differs from Model D, to accurately assess the extent of improvement with this reflection model, we replace the \texttt{reflionxhc} component with the \texttt{cutoffpl} component, and similarly fix the previously mentioned quantities. This model provides a $\chi^2/$d.o.f of 475/373, indicating that our reflection Model D is marginally significant (F-test probability of improvement $\sim8.6\times10^{-3}$) compared to the continuum model.

The ionization parameter ($\log (\xi/\rm erg\ cm\ s^{-1})>4.10$) and the iron abundance ($4.38^{+3.90}_{-2.30}A_{\rm Fe,solar}$) are both found to be significantly high for this observation. Since the super-solar iron abundance likely suggests that the density of the disc ($n_{\rm e}$) is higher than the value assumed in standard reflection models like \texttt{reflionx}, \texttt{relxill} \citep{garcia2016,tomsick2018}, we employ the \texttt{reflionx} based high-density component, \texttt{reflionxhd} \citep{tomsick2018,chakraborty2021}, instead of \texttt{reflionxhc}. Here, $n_{\rm e}$ is a free parameter and can take values ranging from $10^{+15}\ \rm cm^{-3}$ to $10^{+22}\ \rm cm^{-3}$, unlike the previous case where $n_{\rm e}=10^{+15}\ \rm cm^{-3}$ is fixed by default. But, in this model, the illuminating hard coronal emission is described by a power-law with the high energy cutoff fixed at 300 keV, and $A_{\rm Fe}$ is fixed at solar abundance. This model (hereafter referred to as Model E) yields a $\chi^2/$d.o.f of 459.3/371, which is equally good compared to that of Model D. The best-fit parameters (along with their errors) of this model are provided in Table~\ref{tab:astrosat}. The density of the disc ($9.26_{-5.96}^{+16.54}\times10^{+18}\ \rm cm^{-3}$) is indeed found to be high for this observation. The value of the ionization parameter ($\log (\xi/\rm erg\ cm\ s^{-1})=3.78\pm0.15$) is also significantly reduced and well constrained in the present model, as observed in higher-density models previously \citep{tomsick2018,chakraborty2021}.

Thus, we find that both the ionization parameter and disc density of this source are quite high, and such a strongly ionized and dense disc can, in principle, suppress the reflection features \citep{homan2018}. The value of the cross-normalization factor between SXT and LAXPC is found to be in the range of $\sim 1.16-1.27$ in both of these models, which is within the accepted limit \citep{antia2021}.
\subsection{Analysis of the \textit{NuSTAR} spectra}\label{sec:nustar}
We model the 3.0-50.0 keV \textit{NuSTAR} continuum with a thermal BL emission component, \texttt{bbody} and a power-law with a high-energy cutoff, \texttt{cutoffpl}. We fix the neutral hydrogen column density ($N_{\rm H}$) to $2.2\times10^{+22}\ \rm atoms\ cm^{-2}$, as earlier done in the previous sections. Fitting this model to the data yields a $\chi^2/$d.o.f of 724/627. We detect a broad Fe-K$\alpha$ emission line in the spectrum. This feature was also observed by \cite{ludlam2019}, who considered the same data for reflection analysis (see Fig. 1 of \cite{ludlam2019}). To represent this discrete reflection feature, we first add a Gaussian line, \texttt{gauss}, (hereafter referred to as Model F) to our continuum model. This results in a $\chi^2/$d.o.f of 677/624, which corresponds to a $\sim5.8\sigma$ improvement in the spectral fit. The equivalent width of this line is found to be $98_{-55}^{+13}$ eV. The 0.1-50.0 keV unabsorbed flux corresponding to Model F is approximately $\simeq8.5\times10^{-10}\ \rm erg\ cm^{-2}\ s^{-1}$, which is significantly smaller than the fluxes in our sample of \textit{NICER} observations. Only a small fraction of the total flux is contributed by the \texttt{bbody} component (\texttt{bbody} flux is $\simeq0.75\times10^{-10}\ \rm erg\ cm^{-2}\ s^{-1}$). Furthermore, while the blackbody (BB) temperature in this observation is comparable to that in the \textit{NICER} banana state observations, the BB normalization is significantly smaller than that observed in the \textit{NICER} banana state data. It's worth noting that the limited low-energy pass-band of \textit{NuSTAR} restricts our ability to accurately estimate the parameters and flux of the thermal component, which likely contributes to the lower BB normalization and the overall lower luminosity of this observation compared to all the other observations used in this study. Based on the dominance of the Comptonized component, the source was probably in the intermediate or hard state during the \textit{NuSTAR} observation.

\begin{figure}
    \includegraphics[width=0.4\textwidth]{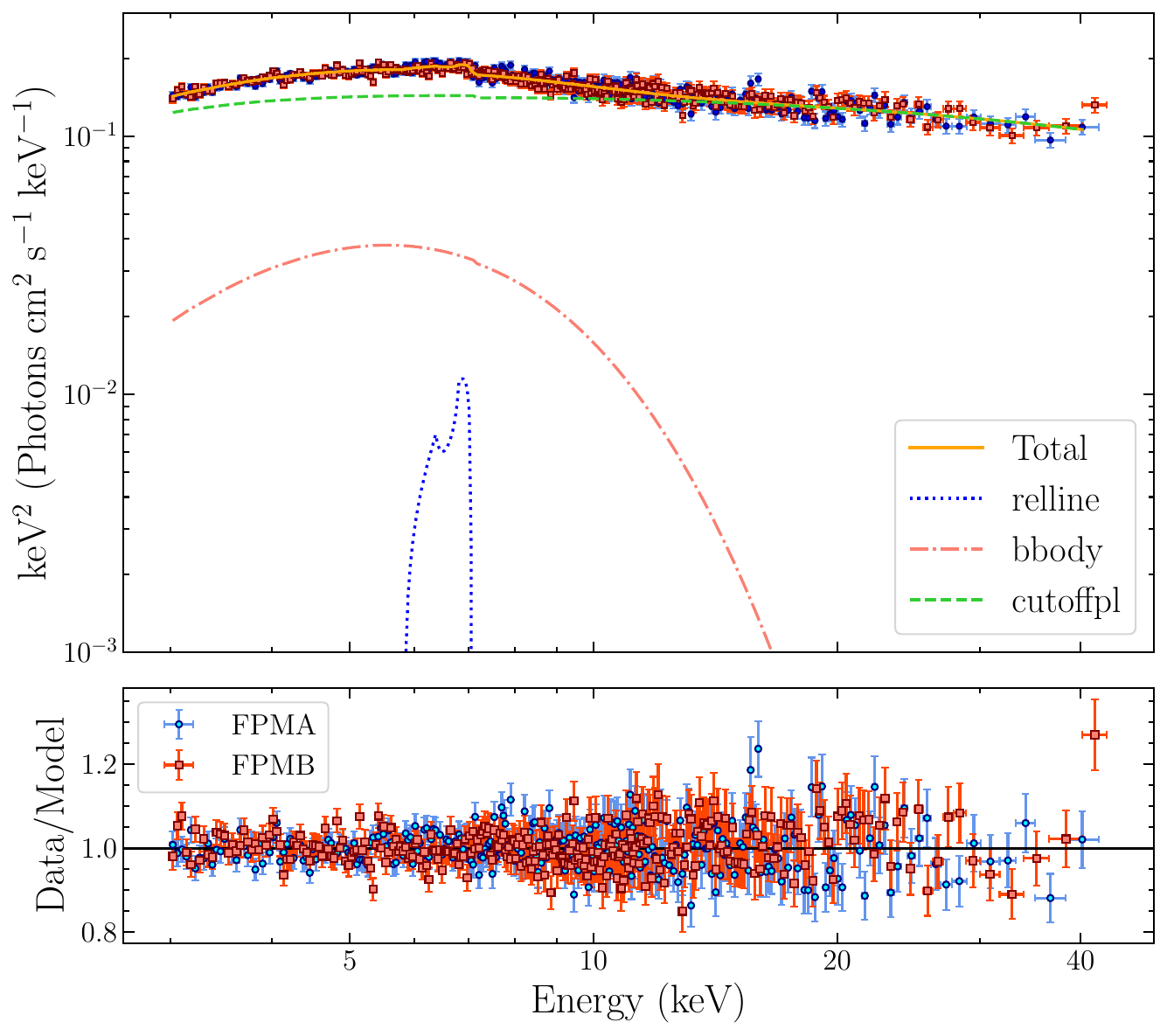}
    \includegraphics[width=0.4\textwidth]{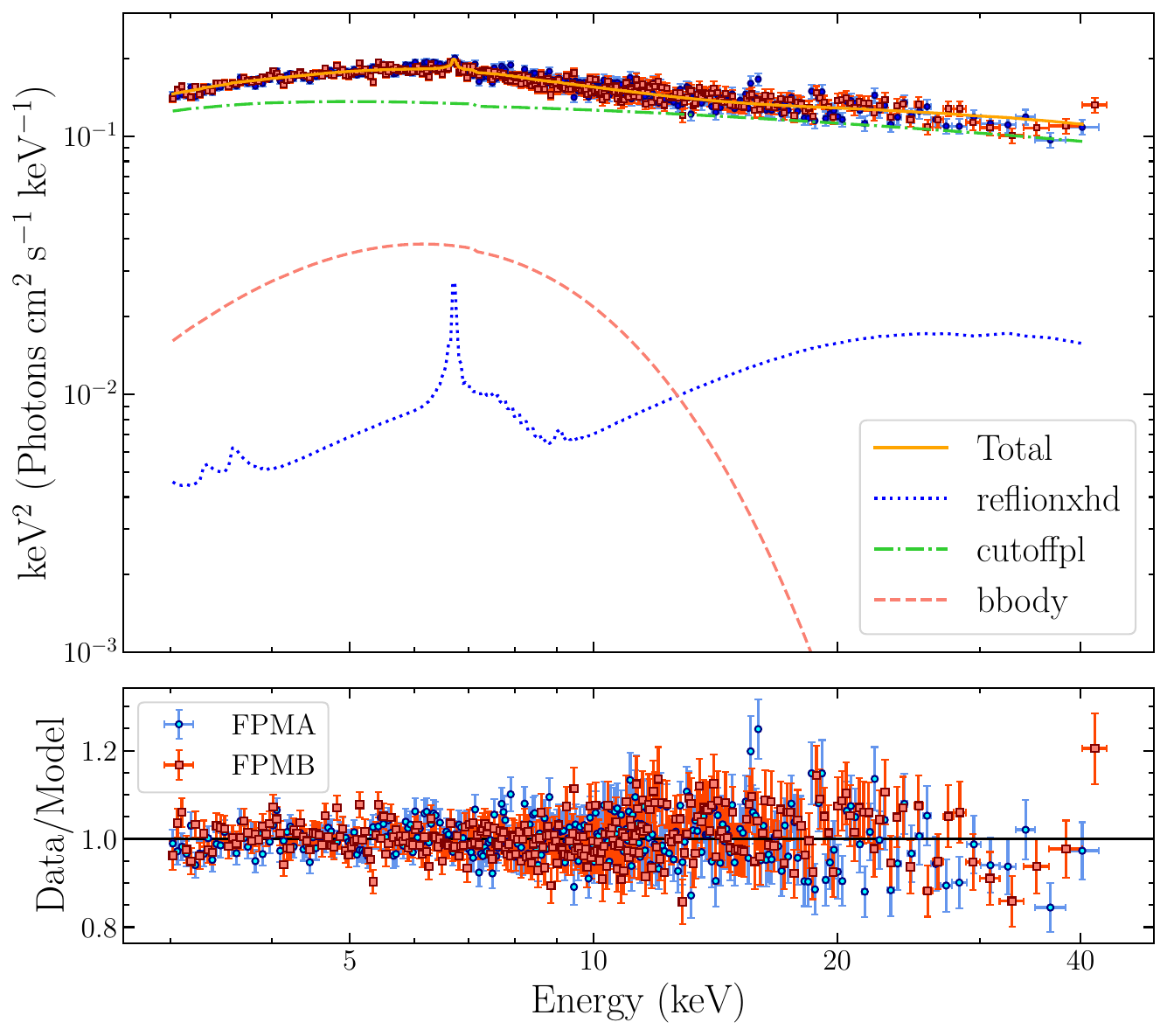}
    \caption{Unfolded spectra with different model components, and ratio of the 3.0-50.0 keV data to the models, Model G (upper panel) and Model H (lower panel), for the \textit{NuSTAR} observation. Data are rebinned for plotting purpose. See Section \ref{sec:nustar} for more details.\label{fig:nuunfold}}
\end{figure}

To fit the broad Fe-K$\alpha$ emission line with a more realistic model, we use a relativistic emission line component, \texttt{relline}, (hereafter referred to as Model G) from the \texttt{relxill} suite. We obtain a slightly worse fit, $\chi^2/$d.o.f of 681.3/624, which may suggest that the emission line is non-relativistic in nature. Additionally, the disc's inner radius from the spectral fit with this model supports this idea: $\mathcal{R}=7.4_{-5.7}^{+18.5}R_{\rm ISCO}$ (where $R_{\rm ISCO}$ is the innermost stable circular orbit radius or ISCO radius). In Model G, we have fixed the inclination $\theta$ and the dimensionless Kerr parameter $a$ to $38^{\circ}$ and 0, respectively, as per \citep{mazzola2019}, since they could not be properly constrained from the spectral fit (the spectral fit remains almost unaltered if we keep $\theta$ as a free parameter). We present our results (Model F and Model G) in Table~\ref{tab:nustar}. The unfolded spectra and the residuals corresponding to Model G are depicted in Fig.~\ref{fig:nuunfold}. In all the models above, a disc component is not significantly detected in this observation, most likely due to \textit{NuSTAR}'s lack of coverage at energies $\leq3.0$ keV. The value of the cross-normalization constant between FPMA and FPMB is found to be $\sim 0.99$ for all the models.

For a detailed investigation of the broad-band continuum and the reflection spectrum, we will now utilise self-consistent reflection models from the \texttt{relxill} \citep{garcia2013,garcia2014} and \texttt{reflionx} \citep{ross2005,ross2007} distribution of models. We first employ the standard version of the reflection model, \texttt{relxill}, where the input illuminating continuum is described by a cutoff power-law component, as previously considered in \cite{ludlam2019} for this observation. The \texttt{relxill} component provides an angle-dependent relativistically smeared reflection spectrum from an illuminated accretion disc along with the input continuum. In this model, we assume a Newtonian emissivity profile, i.e., $q=3$ (setting both the inner and outer emissivity indices to 3), fix the dimensionless Kerr parameter $a$ to 0, and set the outer radius to $400R_g$ (where $R_g=GM/c^2$ is the gravitational radius, with $G$ being the Newtonian gravitational constant, $M$ as the mass of the source, and $c$ as the speed of light in a vacuum).
We keep $\Gamma$, $E_{\rm c}$, $A_{\rm Fe}$, $\log \xi$, inner disc inclination $\theta$, the reflection fraction $f_{\rm refl}$, the disc inner radius $\mathcal{R}_{\rm in}$, and the normalization as free parameters in the model. This model gives a significantly worse fit (compared to the previous models), with a $\chi^2/$d.o.f of 770/624. Besides, as also noted in \cite{ludlam2019}, we obtain an extreme super-solar iron abundance ($\geq 8.92A_{\rm Fe,solar}$) and a high ionization parameter ($\log (\xi/\rm erg\ cm\ s^{-1})=4.11\pm0.13$) in this model. The disc, on the other hand, remains close to the ISCO radius, $\mathcal{R}_{\rm in}=1.46_{-0.46p}^{+0.47}R_{\rm ISCO}$ ($^p$ indicates the parameter pegged at its limit).
The addition of a BL emission component (\texttt{bbody}) to this model significantly improves the fit, providing a $\chi^2/$d.o.f of 675.1/622. Although the values of $\log \xi$ and $A_{\rm Fe}$ remain similarly high ($\log (\xi/\rm erg\ cm\ s^{-1})>3.5$ and $A_{\rm Fe}>1.88A_{\rm Fe,solar}$; the upper limit of both of them gets pegged at the model maximum value) in this model, the disc is found to be truncated, $\mathcal{R}_{\rm in}=22.85_{-9.78}^{+27.15p}R_{\rm ISCO}$, in contrast to the previous model.

Since our previous models suggest that the disc could be truncated and the reflection features are of non-relativistic nature, we replace \texttt{relxill} with the non-relativistic version \texttt{xillver} in our previous model. We maintain identical assumptions about the spectral parameters, except for fixing the inclination angle to $38^{\circ}$ \citep{mazzola2019}. This model gives a slightly better fit with a $\chi^2/$d.o.f of 685.3/624 (leaving $\theta$ as a free parameter in this model gives a $\chi^2/$d.o.f of 685.1/623). The values of both $\log \xi$ and $A_{\rm Fe}$ have been reduced to $\log (\xi/\rm erg\ cm\ s^{-1})=3.66\pm0.15$ and $A_{\rm Fe}=4.61_{-0.94}^{+1.47}A_{\rm Fe,solar}$. However, both the ionization parameter and $A_{\rm Fe}$ remain quite high. In particular, such a high value of $\log \xi$ for a truncated disc is difficult to explain.
We thus consider a high-density disc reflection component \texttt{reflionxhd} from the \texttt{reflionx} suite for the reflection spectrum while keeping the continuum description identical to that of Model F. We will refer to this new model as Model H. We obtain a similarly good fit with a $\chi^2/$d.o.f of 693.1/625. The results are presented in Table~\ref{tab:nustar}. The unfolded spectra and the residuals are depicted in Fig.~\ref{fig:nuunfold}. We find the disc density to be high, $\leq5.3\times10^{19}\ \rm cm^{-3}$, which is consistent with the value obtained previously from the \textit{AstroSat} spectral analysis. Additionally, the ionization parameter ($2.53\pm0.42$) takes a physically consistent value in this model. We further convolve this component with the \texttt{relxill} based relativistic blurring kernel \texttt{relconv} \citep{dauser2010} to check whether our assumption of distant reflection is correct. We find such an addition to be statistically insignificant.

Thus, both the statistical significance and the consistency of physical parameters suggest that the disc could be truncated away from the NS. Additionally, a higher-density disc ($>10^{+15}\ \rm cm^{-3}$) offers an alternative explanation for the extreme super-solar abundance and high ionization parameter as inferred from the \textit{NuSTAR} spectral fit of 4U 1702-429 (obtained with fixed density models).
\section{Discussion}\label{discuss}
In this work, we study the evolution of the inner accretion geometry of the atoll NS-LMXB 4U 1702-429 across different spectral states by performing spectral analyses of 14 \textit{NICER} observations, one \textit{NuSTAR} observation, and one \textit{AstroSat} observation. The 14 \textit{NICER} observations are carefully chosen to cover almost all classical spectral states present in the entire \textit{NICER} sample (see Figures~\ref{fig:hard} and \ref{fig:mhard} and Table \ref{tab:obs1}). We adopt a hybrid model to analyse all the \textit{NICER} observations. For the lower banana (LB) state observations (obs. 3-8), we use a three-component model comprised of two thermal components (disc and BL emission components) and a thermal Comptonization component. For the island/intermediate state observations (obs. 9-14), a two-component model containing a disk emission component and a thermal Comptonization component is used to describe the spectral properties. However, we could not find a suitable statistically significant Comptonization component for one of the upper banana (UB) state observations. In the other UB state observation (obs. 2), we represent the hard X-ray emission using a pegged power-law component. Weak Fe-K$\alpha$ emission lines (with a significance of $\sim 3-4\sigma$) are only detected in observations 10 and 11. The estimated values of the soft spectral parameters, associated with the disc and BL components, fall within the typical range of such parameters observed in other NS-LMXBs \citep{lin2007,wang2019,padilla2017,ludlam2020}, and are also in agreement with previous studies of 4U 1702-429 \citep{iaria2016,mazzola2019}.

\subsection{Evolution of disc and boundary layer parameters across different spectral states}
The unabsorbed flux in our selected \textit{NICER} observations (0.1-50.0 keV) ranges from approximately $12.5\times10^{-10}$ to $23.0\times10^{-10}\ \rm erg\ cm^{-2}\ s^{-1}$, with banana state observations generally more luminous than island/intermediate state observations (see Table~\ref{tab:nicerflux}). Assuming a 5.5 kpc distance \citep{iaria2016}, this translates to a luminosity range of approximately $0.025-0.05L_{\rm Edd}$. In the UB state, the disc and BL contribute over $80\%$ of X-ray photons, decreasing to $50-60\%$ in the LB state. The BL flux fraction drops from $50\%$ to about $20\%$, while the disc flux fraction decreases from $50\%$ to approximately $40\%$. However, the variation in the disc and BL components within these two states is minor. Therefore, the evolution of the system in the banana state is primarily controlled by the BL component, with the UB to LB transition likely caused by a change in BL geometry. As the source shifts to the island/intermediate state, $kT_{\rm in}$ decreases, and the disc inner radius increases, indicating the disc moving farther from the NS. The thermal flux (disc+BL flux) fraction drops below $40\%$, making Comptonized emission dominant.  Additionally, the presence of the BL component is only indirectly detected as a source of the soft seed photons for Comptonization. Beyond observation 7, $kT_{\rm in}$ drops from 0.55-0.6 keV to below 0.35 keV, and the disc recedes beyond $\sim60$ km, causing a significant decrease in accretion rate, and subsequently, in BL flux and temperature \citep{frank2002,padilla2017}.

The observed spectral parameter evolution aligns with trends in other atoll NS-LMXBs like 4U 1608-428 \citep{padilla2017}. This accretion geometry mirrors that of BH-LMXBs, with the NS/BL surface contributing additional soft X-ray photons. It is important to note that the estimated Comptonized flux in the island/intermediate state may be a lower limit due to \textit{NICER}'s limited high-energy coverage. Furthermore, the high value of $N_{\rm H}$ can affect the computation of the spectral parameters associated with the thermal components (disc/BL) for the island/intermediate observations. Additionally, the restricted energy range of the \textit{NICER} data makes it challenging to precisely determine the spectral state of observations 9-14 (and observation 8), leading to the categorisation of all of them as island/intermediate state observations.

\subsection{A comparative study of our findings and a previous work on 4U 1702-429}
In our study we find that the disc/BB temperature in the banana state is higher than that in the island/intermediate state, which is contrary to the trend mentioned in \cite{mazzola2019}. In their work, \cite{mazzola2019} employed a two component model (which corresponds to our Model C) to describe the 0.3-60 keV continuum of a \textit{XMM-Newton}/\textit{INTEGRAL} observation. They also applied this model to three \textit{BeppoSAX} observations, considering the energy range of 0.1-100 keV. Their analysis revealed that the unabsorbed luminosity in the 0.1-100.0 keV range for the \textit{XMM-Newton}/\textit{INTEGRAL} observation was approximately 2 to 3 times higher than that for the three  \textit{BeppoSAX} observations. Additionally, they estimated the electron temperature $kT_{\rm e}$ in the \textit{XMM-Newton}/\textit{INTEGRAL} observation to be significantly lower than that  in the three  \textit{BeppoSAX} observations. Based on these two findings, \cite{mazzola2019} categorised the  \textit{XMM-Newton}/ \textit{INTEGRAL} observation as a soft state observation while classifying other three observations as hard state observations. However, there are certain conceptual issues with their assertion that call its reliability into question.  Firstly, the BB temperature in the  \textit{XMM-Newton}/ \textit{INTEGRAL} observation is lower than that in the three  \textit{BeppoSAX} observations, and $kT_{\rm in}$ is similar (infact, slightly higher in first two \textit{BeppoSAX} observations) in all these observations. Besides, the disc is much closer to the NS surface in the three \textit{BeppoSAX} observations than in the \textit{XMM-Newton}/ \textit{INTEGRAL} observation. Most notably, the Comptonized flux in the 0.1-100.0 keV energy range is observed to be nearly twice as high as the soft component flux (i.e., disc flux as they do not consider a BL emission component in their model) in the \textit{XMM-Newton}/\textit{INTEGRAL} spectra. These findings are fundamentally at odds with our current understanding of the accretion process in atoll NS-LMXBs. As a result, the claim made by \cite{mazzola2019} regarding the spectral state classification of these observations poses significant challenges in terms of explanation and interpretation. 

\subsection{Investigating the reasons behind the weakness/absence of reflection features}
In all the \textit{NICER} observations used in this work, the reflection features are either absent or faint. To investigate this issue further, we analyse broad-band \textit{NuSTAR} and \textit{AstroSat} spectra in the energy range 3.0-50.0 keV, and 1.3-25 keV, respectively. Comparing the values of disc temperature, BB temperature, and soft thermal (BB+disc) flux fraction with our \textit{NICER} analyses suggests that the source was in the banana state (likely in the LB state) during the \textit{AstroSat} observation (see first paragraph of Section \ref{sec:astrosat} for more details). We do not detect reflection features in this observation, similar to the \textit{NICER} banana state observations. 
Our analyses with reflection models for this observation leads to a high value of the ionization parameter $\log (\xi/\rm erg\ cm\ s^{-1})=3.78\pm0.15$ and disc density $n_{\rm e}=9.26_{-5.96}^{+16.54}\times10^{+18}\ \rm cm^{-3}$ when the iron abundance $A_{\rm Fe}=A_{\rm Fe,solar}$ (provides an even higher value of  $\log \xi$ when $A_{\rm Fe}$ is a free parameter in the model). A higher value of $\log (\xi/\rm erg\ cm\ s^{-1})\sim3.8$ results in a highly ionized disc atmosphere to a greater optical depth (due to a hotter disc atmosphere), causing the Fe K$\alpha$ line emission to become smeared out \citep{garcia2013}. On the other hand, at such high densities, free-free heating is also enhanced, raising the temperature of the disc atmosphere, which in turn weakens and broadens the spectral features through Compton scattering \citep{garcia2022}. So, both these effects together make the reflection features hard to be detected. 
In the banana state, the disc stays close to the NS surface or BL (see Table~\ref{tab:nicer} and \ref{tab:astrosat}, and Fig.~\ref{fig:spec}). Since the BL could play an important role in serving as the primary continuum for the reflection \citep{cackett2010,garcia2022}, such a large value of $\log \xi$ in the banana state is perhaps not unexpected. Besides, a simple corona-disk model of \cite{zdziarski1994} indicates a higher disc density for NSs than BHs (unless the accretion rate is not very low) \citep{garcia2022}. Recently, disc densities greater than $10^{15}\ \rm cm^{-3}$ have been reported for several NS-LMXBs, such as Cyg X-2 \citep{ludlam2022} and 4U 1735-44 \citep{ludlam2020}. Although \citet{mazzola2019} used a fixed-density reflection model (i.e., $n_{\rm e} = 10^{15}\ \rm cm^{-3}$) in their work on 4U 1702-429, they calculated the value of $n_{\rm e}$ from the standard relation between $\xi$ and $n_{\rm e}$ and found it to be $\sim10^{21}\ \rm cm^{-3}$ for the \textit{XMM-Newton}/\textit{INTEGRAL} observation. Therefore, the possibility of a higher-density disc for 4U 1702-429 is also reasonable.

The \textit{NuSTAR} observation, on the other hand, is less luminous (0.1-50 keV unabsorbed X-ray flux is $\simeq8.5\times10^{-10}\ \rm erg\ cm^{-2}\ s^{-1}$) than all the observations considered in this work. The presence of a disc is not detected in the spectra, and only a small fraction of the flux ($\sim9\%$) is found to be contributed by the BL. Here, we note that the unavailability of data below 3 keV restricts any proper estimation of the soft X-ray flux, thus making a reliable determination of the state of this observation challenging through spectral analysis alone. Unlike the \textit{AstroSat} observation, we observe a weak iron emission line in the \textit{NuSTAR} observation. In our \textit{NICER} sample, we only detect this reflection feature significantly in two island/intermediate state observations (obs. 10 and 11), but the presence of this line is less significant ($\sim3-4\sigma$) in these two observations compared to the \textit{NuSTAR} observation ($\sim5.8\sigma$). 
This emission line was detected with a very high significance in the \textit{XMM-Newton} data \citep{iaria2016,mazzola2019}, where the F-test probability of chance improvement was $\sim10^{-30}$ when a Gaussian line was added to the continuum \citep{iaria2016}.
Our analyses with the relativistic emission line model \texttt{relline} and reflection models \texttt{relxill} suggest that the disc could be truncated ($\mathcal{R}_{\rm in}\gtrsim 2R_{\rm ISCO}$). But, both the $\log \xi$ and $A_{\rm Fe}$ remain significantly high in our analysis with \texttt{relxill} ($\log (\xi/\rm erg\ cm\ s^{-1})>3.5$ and $A_{\rm Fe}>1.9A_{\rm Fe,solar}$). A large value of $\log \xi$ and the disc truncation both result in a weak iron emission line \citep{garcia2013,homan2018}. Also, super-solar abundance (and high ionization) can lead to a higher density disc as discussed above.  Motivated by our analyses with \textit{AstroSat} data, we also employ a high density disk distant reflection model, \texttt{reflionxhd} \citep{tomsick2018}, and find the disc to have a moderate ionization  ($2.53\pm0.42$) and high density ($>10^{15}\ \rm cm^{-3}$). Notably, the $\log \xi$ value in this model was slightly smaller than that estimated for the \textit{XMM-Newton}/INTEGRAL observation ($3.0_{-0.3}^{+0.1}$), but roughly similar to that obtained for the \textit{BeppoSAX} observations ($\sim2.5$) \citep{mazzola2019}.
We also notice that the above fit does not improve when we convolve a relativistic blurring kernal with \texttt{reflionxhd}. Besides, a Gaussian line provides a slightly better description of the emission line than \texttt{relline}. Hence, the Fe-K$\alpha$ emission line is likely of non-relativistic nature  and originates in a disc situated away from the NS surface. 
\section{Conclusion}\label{conc}
In this work, we investigate the evolution of inner accretion geometry of the atoll NS-LMXB 4U 1702-429 across various spectral states using 14 \textit{NICER} observations, one \textit{NuSTAR} observation, and one \textit{AstroSat} observation. Although broad-band analyses of this source have been performed with \textit{XMM-Newton}, \textit{INTEGRAL}, \textit{BeppoSAX}, and \textit{NuSTAR}, there were several inconsistencies in earlier works, primarily related to the extent of inner disc radius and disc/BL parameters in different spectral states (discussed in detail in Section \ref{discuss}). We summarize our main results as follows.
\begin{itemize}
    \item We require a hybrid model to analyse 14 \textit{NICER} observations, spanning a range in Eddington Luminosity of approximately $0.025-0.05L_{\rm Edd}$. Our model for banana state observations is composed of two thermal components, originating from the disc and BL, and a hard Comptonized component, while the BL component is not directly detected in the island/intermediate state observations. However, the BL serves as a primary source of soft seed photons in all these observations.
    \item We find that the transition from UB to LB state is associated with a substantial decrease in the BL flux fraction, suggesting that changes in the geometry of the BL play a pivotal role in this evolution. In the island/intermediate state, the disc moves farther from the NS, leading to a decrease in disc temperature and inner radius. Comptonized emission becomes dominant, marking a significant shift in the accretion process.
    \item  The absence or faintness of reflection features in \textit{NICER} observations prompts a thorough investigation using \textit{NuSTAR} and \textit{AstroSat} broad-band data. In the \textit{AstroSat} observation, a highly ionized disc is suggested as a plausible explanation for the absence of reflection features. The \textit{NuSTAR} observation, less luminous than \textit{NICER} and \textit{AstroSat} observations, exhibits a weak iron emission line, suggesting a non-relativistic origin from a truncated disc away from the NS surface.
\end{itemize}
\section*{Acknowledgements}
We express our gratitude to the anonymous referee for his/her constructive comments, which contributed to the improvement of the manuscript. This research utilized data, software, and web tools obtained from the High Energy Astrophysics Science Archive Research Center (HEASARC), a service of the Astrophysics Science Division at NASA/GSFC. We acknowledge the use of data from the \textit{AstroSat} mission of the Indian Space Research Organisation (ISRO), archived at the Indian Space Science Data Centre (ISSDC). Special thanks to the POC teams of the SXT and LAXPC instruments for archiving data and providing essential software tools. This research has also made use of NuSTAR Data Analysis Software (NuSTARDAS) jointly developed by the ASI Science Data Center (ASDC, Italy) and the California Institute of Technology (Caltech, USA). We extend our appreciation to John Tomsick for providing the reflionxhd model, and we thank Michael Parker for developing this model. S. Banerjee would like to express gratitude to Gulab Chand Dewangan for his valuable suggestions. Although our work has diverged significantly from its original goal, we wish to thank Sudip Bhattacharyya for initiating this project. S. Banerjee is also grateful to him for his continuous support and encouragement throughout the course of this work.
\section*{Data Availability}
The  observational  data  used  in  this  paper  are  publicly available at HEASARC (\url{https://heasarc.gsfc.nasa.gov/}) and AstroSat data archive (\url{https://astrobrowse.issdc.gov.in/astro_archive/archive/Home.jsp}). Any  additional  information  will  be  available upon reasonable request.


\bibliographystyle{mnras}
\bibliography{4u1702} 

\begin{table*}
\renewcommand{\arraystretch}{1.5}
\caption{Results of spectral fitting of the \textit{NICER} observations. The best-fit parameter values and the corresponding errors at 90$\%$ confidence level for the Models A, $\mathscr{A}$, B, C, and $\mathscr{C}$ are presented. In \texttt{XSPEC} notation, Model A: \texttt{tbabs*(diskbb+bbody)},  Model $\mathscr{A}$: \texttt{tbabs*(diskbb+bbody+pegpwrlw)}, Model B: \texttt{tbabs*(diskbb+bbody+nthcomp)}, Model C: \texttt{tbabs*(diskbb+nthcomp)}, and Model $\mathscr{C}$: \texttt{tbabs*(diskbb+nthcomp+gauss)}.\label{tab:nicer}}
  \begin{tabular}{|c|c|c|c|c|c|c|c|c|c|c|c|}
    \hline
     Obs ID & Model &  $kT_{\rm in}$ & $N_{\rm diskbb}$ (km) & $kT_{\rm BB}$ (keV) & $N_{\rm bbody}\ (\rm erg\ s^{-1})$  & $\Gamma$ & $kT_{\rm e}$  & $N_{\rm comp}$ & $\chi^2/\rm d.o.f$\\
              &   & (keV)  &  ($R_{\rm in}\sqrt{\cos \theta})$ & (keV) & ($L_{36}/D_{10}^2$) & & (keV) \\
    \hline
         1050110113 & A & $0.75\pm0.02$ & $7.28\pm0.34$ & $1.86\pm0.02$  & $13.29\pm0.36$ & - & - & - &346.1/303\\
         1050110131 & $\mathscr{A}$ & $0.72\pm0.04$  & $6.32\pm0.82$ & $1.72\pm0.04$ & $11.26\pm1.07$& $2.86_{-0.50}^{+0.14p}$ & - & $97.46_{-75.03}^{+213.65}$ &670.8/572\\
         1050110129 & B & $0.61\pm0.03$ & $9.74_{-0.89}^{+1.15}$ & $1.34\pm0.16$ & $4.05\pm2.17$ & $2.0^f$ & $10^f$ & $9.13\pm5.83\times10^{-3}$ &498.4/475\\
         1050110130 & B & $0.57\pm0.04$  & $10.75_{-1.22}^{+1.61}$ & $1.24\pm0.16$ & $3.64\pm1.52$ & $2.0^f$ & $10^f$ & $9.67\pm5.31 \times10^{-3}$ &334.8/293\\
         1050110128 & B & $0.54\pm0.05$  &  $11.95_{-1.65}^{+2.70}$ & $1.13\pm0.20$ & $2.51\pm1.66$ & $2.0^f$ & $10^f$ & $13.41\pm9.05\times10^{-3}$ &347.3/309\\
         1050110118 & B & $0.56\pm0.03$ & $10.01_{-0.83}^{+0.98}$ & $1.30\pm0.16$ & $4.44\pm1.97$ & $2.0^f$ & $10^f$ & $6.56\pm4.77 \times10^{-3}$ &301.3/254\\
         1050110126 & B & $0.58\pm0.04$  & $8.86_{-0.88}^{+1.21}$ & $1.24\pm0.24$ & $2.49\pm2.10$ & $2.0^f$ & $10^f$ & $7.92\pm7.24 \times10^{-3}$ &269.2/226\\
         1050110117 & B & $0.32\pm0.03$  & $31.03_{-7.36}^{+13.27}$ & $0.67\pm0.13$ & $3.87\pm1.52$ & $<1.79$ & $1.94_{-0.15}^{+1.24}$ & $7.49_{-5.17}^{+72.72} \times10^{-3}$ &373.2/380\\
         2587030103 & C & $0.25\pm0.01$  & $37.44^{+6.00}_{-5.24}$ & $0.45\pm0.02$ & - & $1.76\pm0.02$ & $3.13\pm0.13$& $84.51\pm4.25\times10^{-3}$ &702.6/602\\
         2587030101 & $\mathscr{C}$ & $0.23\pm0.02$  & $50.65_{-10.56}^{+12.65}$  & $0.47\pm0.03$ & - & $1.81\pm0.04$& $3.31\pm0.42$& $78.63\pm5.05\times10^{-3}$ & 602.4/518\\
         2587030104 & $\mathscr{C}$ & $0.24\pm0.02$  & $37.85_{-8.48}^{+10.23}$ & $0.50\pm0.04$ & - &$1.81\pm0.08$&$>3.25$& $74.66\pm5.14\times10^{-3}$ &594.7/577\\
         0050110105 & C &  $0.21\pm0.06$  & $33.98_{-18.85}^{+46.32}$ & $0.50\pm0.05$ & - & $1.79\pm0.06$ & $2.95_{-0.41}^{+1.00}$ & $68.44\pm5.07\times10^{-3}$ &443.7/377\\
         0050110107 & C &  $0.27\pm0.06$  & $18.38_{-7.41}^{+15.08}$ & $0.59\pm0.06$& - & $1.92\pm0.08$ & $>3.25$ & $58.46\pm4.43\times10^{-3}$ & 347.6/375\\
         0050110110 & C &  $0.17\pm0.04$ & $74.17_{-42.34}^{+103.54}$ & $0.49\pm0.03$ & - & $1.72\pm0.05$ & $2.51\pm0.26$& $69.99\pm3.20\times10^{-3}$ &514.0/449\\
    \hline
\end{tabular}
\begin{flushleft}
\textbf{Notes.} $^p$ and $^f$ denote that the parameter is pegged at its limit and fixed during the fit, respectively. $N_{\rm comp}$ refers to the \texttt{nthcomp} normalization for all the observations except 1050110131, for which this quantity gives the normalization of the \texttt{pegpwrlw} component. See Section \ref{sec:nicer} for more details about these models.
\end{flushleft}
\end{table*}

\begin{table*}
\renewcommand{\arraystretch}{1.3}
\caption{Unabsorbed flux value of different components in the energy band 0.1-50.0 keV of the \textit{NICER} observations. The flux values are given in the unit of $10^{-10}\ \rm erg\ cm^{-2}\ s^{-1}$.}\label{tab:nicerflux}
  \begin{tabular}{|c|c|c|c|}
    \hline
     Obs ID & Disc Flux & BB Flux & Total Flux\\
    \hline
         1050110113 & $11.89\pm0.17$ & $11.07\pm0.30$ & $22.96\pm0.31$\\
         1050110131 & $7.61\pm0.46$  & $9.41\pm0.09$  & $17.85\pm0.48$\\
         1050110129 & $9.14\pm0.15$  & $3.40\pm0.09$ & $19.46\pm0.15$\\
         1050110130 & $8.72\pm0.22$  & $3.05\pm0.10$  & $19.06\pm0.17$\\
         1050110128 & $8.44\pm0.30$  & $2.09\pm0.14$  & $18.96\pm0.20$\\
         1050110118 & $6.97\pm0.09$  & $3.72\pm0.10$ & $16.14\pm0.20$\\
         1050110126 & $6.10\pm0.08$  & $2.09\pm0.11$  & $14.18\pm0.20$\\
         1050110117 & $6.73\pm0.23$  & $3.25\pm0.10$  & $16.33\pm0.32$\\
         2587030103 & $3.62\pm0.14$  & -  & $16.49\pm0.19$\\
         2587030101 & $4.66\pm0.24$  & -  & $16.77\pm0.26$\\
         2587030104 & $3.10\pm0.19$  & -  & $15.81\pm0.37$\\
         0050110105 & $1.48\pm0.57$  &  -  & $12.64\pm0.56$\\
         0050110107 & $1.28\pm0.24$  &  -  & $13.50\pm0.81$\\
         0050110110 & $2.55\pm1.37$  &  - & $13.99\pm0.44$\\
    \hline
 \end{tabular}
 \end{table*}

\begin{table*}
\caption{Results of spectral fitting of the \textit{AstroSat} observations. Best-fit parameter values and the corresponding errors at 90$\%$ confidence level for the Models B, D and E. In \texttt{XSPEC} notation, Model B: \texttt{tbabs*con*(diskbb+bbody+nthcomp)}, Model D: \texttt{tbabs*con*(diskbb+bbody+reflionxhc)} and Model E: \texttt{tbabs*con*(diskbb+bbody+reflionxhd)}.\label{tab:astrosat}}
\centering
\begin{tabular}{clccccc}
\hline 
Spectral Components & Parameters & Model~B & Model ~D & Model ~E\\ 
\hline
{\sc Constant}& C$_{\rm SXT}$&$1.0^f$&$1.0^f$&$1.0^f$\\

&C$_{\rm LAXPC}$&$1.23\pm0.04$&$1.21\pm0.05$&$1.23\pm0.04$\\

{\sc tbabs} &$N\rm_{H}~(10^{22}\ \rm atoms\ cm^{-2})$&$2.11\pm0.11$&$2.2^f$&$2.2^f$\\

{\sc diskbb} &$kT\rm_{in}$~(keV)&$0.56\pm0.02$&$0.52\pm0.04$ &$0.39\pm0.09$\\

&$N\rm_{\rm diskbb}$  ($R_{\rm in}\sqrt{\cos \theta}$) (km)&$10.94_{-1.21}^{+1.41}$&$8.14_{-1.83}^{+1.65}$ & $11.82_{-5.39}^{+6.16}$\\


{\sc bbody} &$kT\rm_{\rm BB}$&$1.44\pm0.05$&$1.81\pm0.15$&$1.81\pm0.08$\\

&$\rm Norm (L_{36}/D_{10}^2)$ ($\rm erg\ s^{-1}$)&$4.28\pm0.37$&$2.78\pm0.40$&$3.08\pm0.25$\\

{\sc nthcomp} &$\Gamma$&$2.05\pm0.08$\\

&$kT_{\rm e}$ (keV) &$100^f$\\

&$kT\rm_{\rm seed}$&$1.44\pm0.05$&\\

&$\rm Norm(\times10^{-3})$&$4.44\pm0.81$\\

{\sc refionxhc}& $\Gamma$&&$1.83\pm0.07$&\\

&$E_{\rm c}$ (keV)& &$300^f$\\

&$\log (\xi/\rm erg\ cm\ s^{-1})$&&$>4.10$&\\

&$A_{\rm Fe}$(in solar unit)&&$4.38_{-2.30}^{+3.90}$&\\

&$\rm Norm(\times 10^{-7})$&&$1.27\pm0.29$&\\

{\sc reflionxhd} &$\Gamma$&&&$1.75\pm0.07$\\

&$E_{\rm c}$ (keV)& &&$300^f$\\

&$\log (\xi/\rm erg\ cm\ s^{-1})$&&&$3.78\pm0.15$\\

&$A_{\rm Fe}$(in solar unit)&&&$1.0^f$\\

&$n_{\rm e}$ ($10^{+18}\ \rm cm^{-3}$)&&&$9.26_{-5.96}^{+16.54}$\\

&$\rm Norm$&&&$1.15\pm0.26$\\

\hline
$\chi^2/$d.o.f& &482.8/372&462.9/371&459.3/371\\
\hline
\end{tabular}
\begin{flushleft}
{\bf Note:} In this table, $^f$ means that the parameter is fixed during the fit. $\rm Norm$ refers to normalization of the respective model. See Section \ref{sec:astrosat} for more details about these models.
\end{flushleft}
\end{table*}
\begin{table*}
\caption{Results of spectral fitting of the \textit{NuSTAR} observations. Best-fit parameter values and the corresponding errors at 90$\%$ confidence level for the Models F, G, and H. In \texttt{XSPEC} notation, Model F: \texttt{tbabs*con*(bbody+cutoffpl+gauss)}, Model G: \texttt{tbabs*con*(bbody+cutoffpl+relline)}, and Model H: \texttt{tbabs*con*(cutoffpl+bbody+reflionxhd)}.\label{tab:nustar}}
\centering
\begin{tabular}{clccccc}
\hline 
Spectral Components & Parameters & Model F & Model G & Model H\\ 
\hline
{\sc Constant}& C$_{\rm FPMA}$&$1.0^f$&$1.0^f$&$1.0^f$\\

&C$_{\rm FPMB}$&$0.995\pm0.004$&$0.995\pm0.004$&$0.995\pm0.005$\\

{\sc tbabs} &$N\rm_{H}~(10^{22}\ \rm atoms\ cm^{-2})$ &$2.2^f$&$2.2^f$&$2.2^f$\\

{\sc bbody} &$kT\rm_{\rm BB}$&$1.36\pm0.05$&$1.38\pm0.05$&$1.54\pm0.08$\\

&$\rm Norm (L_{36}/D_{10}^2)$ ($\rm erg\ s^{-1}$) &$0.94\pm0.13$&$1.02\pm0.10$&$1.02\pm0.06$\\

{\sc gauss} &$E_{\rm line}$~(keV)&$6.60\pm0.20$\\

&$\sigma$~(keV)&$0.67\pm0.31$\\

&$\rm Norm(\times 10^{-4})$&$3.26\pm1.77$\\

{\sc relline} &$E_{\rm line}$~(keV)&&$6.62\pm0.15$\\

&$\theta$~(degree) &&$38^f$\\

&$\mathcal{R_{\rm in}}$ ($R_{\rm ISCO})$&&$7.39_{-5.69}^{+18.48}$\\

&$\rm Norm(\times 10^{-4})$&&$1.83\pm0.79$\\

{\sc reflionxhd} &$\log (\xi/\rm erg\ cm\ s^{-1})$&&&$2.53\pm0.42$\\

&$A_{\rm Fe}$(in solar unit)&&&$1.0^f$\\

&$n_{\rm e}$ ($\rm cm^{-3}$)&&&$\leq5.30\times10^{+19}$\\

&$\rm Norm$&&&$6.04_{-5.56}^{+6.65}$\\

{\sc cuoffpl} &$\Gamma$&$2.01\pm0.04$&$2.01\pm0.05$&$2.14\pm0.05$\\

&$E_{\rm c}$ (keV) &$105.38_{-25.60}^{+48.80}$&$108.97_{-26.47}^{+65.34}$&$300^f$\\

&$\rm Norm.$&$0.16\pm0.01$&$0.16\pm0.01$&$0.18\pm0.01$\\

\hline
$\chi^2/$d.o.f && 677/624&681.3/624&693.1/625\\
\hline
\end{tabular}
\begin{flushleft}
{\bf Note:} In this table, $^f$ means that the parameter is fixed during the fit. $\rm Norm$ refers to normalization of the respective model. See Section \ref{sec:nustar} for more details about these models.
\end{flushleft}
\end{table*}



\bsp	
\label{lastpage}
\end{document}